\documentclass[lettersize,journal]{IEEEtran}
\usepackage{amsmath,amsfonts}
\usepackage{algorithmic}
\usepackage{algorithm}
\usepackage{array}

\usepackage{textcomp}
\usepackage{stfloats}
\usepackage{url}
\usepackage{verbatim}
\usepackage{graphicx}
\usepackage{cite}
\usepackage{subfigure}
\usepackage{xcolor}

\usepackage{booktabs}

\begin{document}

\title{Resonant Beam Enabled DoA Estimation in  Passive Positioning System}

\author{\normalsize Yixuan Guo, Qingwei Jiang, Mengyuan Xu, Wen Fang, Qingwen Liu,~\IEEEmembership{Senior Member,~IEEE}, Gang Yan,~\IEEEmembership{Member,~IEEE},  Qunhui Yang,~\IEEEmembership{Senior Member,~IEEE}, and Hai Lu
\thanks{
This work was supported in part by the Interdisciplinary
Project in Ocean Research of Tongji University; the National Natural Science Foundation of China under Grant 62071334, Grant 61771344, Grant 62301308, and Grant 62305019; the National Key Research and Development Project under Grant 2022YFA1004700, Grant 2020YFB2103900, and Grant 2020YFB2103902; the Shanghai Municipal Science and Technology Major Project under Grant 2021SHZDZX0100; the Shanghai Municipal Commission of Science and Technology Project under Grant 19511132101; the China National Postdoctoral Program for Innovative Talents under Grant BX20230223; the China Postdoctoral Science Foundation under Grant 2023M732262; the Natural Science Foundation of Shanghai under Grant 22ZR1462900; and the Fundamental Research Funds for the Central Universities under Grant 22120210543.
\textit{(Corresponding authors: Mengyuan Xu; Qingwen Liu.)}

Y. Guo is with the Shanghai Research Institute for Intelligent Autonomous Systems, Tongji University, Shanghai 201210, China 
(e-mail: guoyixuan@tongji.edu.cn).

Q. Jiang, M. Xu, and Q. Liu are with the College of Electronics and Information Engineering, Tongji University, Shanghai 201804, China
(e-mail: jiangqw@tongji.edu.cn, xumy@tongji.edu.cn, qliu@tongji.edu.cn).

W. Fang is with the School of Electronic Information and Electrical Engineering, Shanghai Jiao Tong University, Shanghai 200240, China (email: wendyfang@sjtu.edu.cn).

G. Yan is with the School of Physics Science and Engineering, Tongji University, Shanghai 200092, China (e-mail: gyan@tongji.edu.cn).

Q. Yang is with State Key Laboratory of Marine Geology, and Project Management Office of China National Scientific Seafloor Observatory, Tongji University, Shanghai, 200092, China (e-mail: yangqh@tongji.edu.cn).

H. Lu is with the Project Management Office of China National Scientific Seafloor Observatory, Tongji University, Shanghai, 200092, China (e-mail: lh@tongji.edu.cn).
	}
}

\maketitle

\begin{abstract}
The rapid advancement of the next generation of communications and internet of things (IoT) technologies has made the provision of location-based services for diverse devices an increasingly pressing necessity. \textcolor{black}{Localizing devices with/without intelligent computing abilities, including both active and passive devices is essential}, especially in indoor scenarios. For traditional RF positioning systems, aligning transmission signals and dealing with signal interference in complex environments are inevitable challenges. 
Therefore, this paper proposed a new passive positioning system, the RF-band resonant beam positioning system (RF-RBPS), which \textcolor{black}{achieves energy concentration and beam alignment by amplifying echoes between the base station (BS) and the passive target (PT), without the need for complex channel estimation and time-consuming beamforming and provides high-precision direction of arrival (DoA) estimation for battery-free targets using the resonant mechanism.}
The direction information of the PT is estimated using the multiple signal classification (MUSIC) algorithm at the end of BS. The feasibility of the proposed system is validated through theoretical analysis and simulations. 
Results indicate that the proposed RF-RBPS surpasses RF-band active positioning system (RF-APS) in precision, \textcolor{black}{achieving millimeter-level precision at 2m within an elevation angle of 35$^\circ$, and an error of less than 3cm at 2.5m within an elevation angle of 35$^\circ$.}
\end{abstract}


\section{Introduction}
\begin{figure}[!t]
\centering
\includegraphics[width=3in]{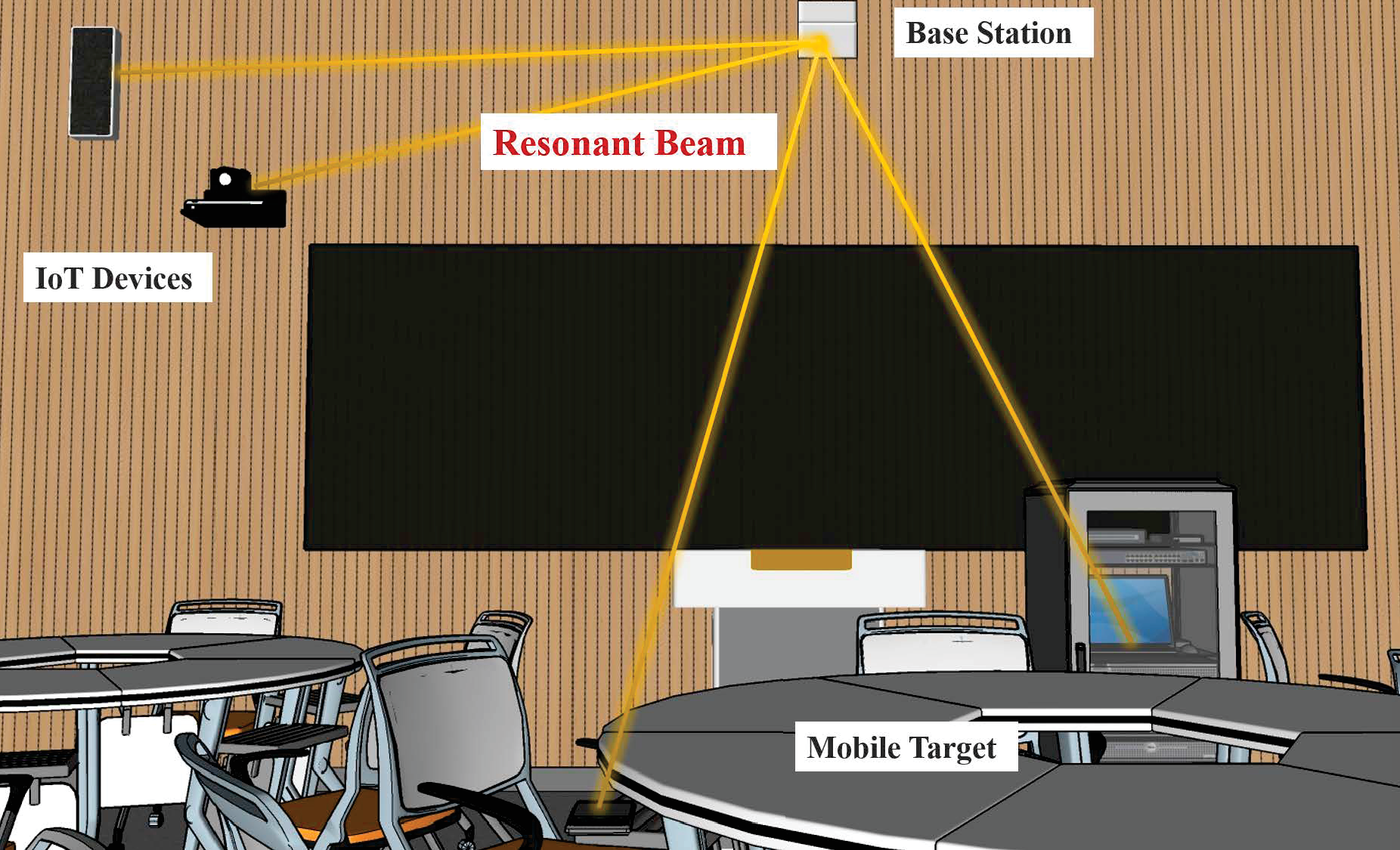}
\caption{A typical application scenario of resonant beam enabled passive positioning system.}
\label{fig_1}
\end{figure}

\IEEEPARstart{I}{n} today's increasingly globalized and digital world, precise positioning technology has become a key component of countless industries and fields. From the internet of things (IoT) device management, to intelligent transportation systems, to augmented reality (AR) and virtual reality (VR) applications, positioning technology is leading a series of revolutionary changes. 
Concurrently, the ubiquity of wireless connectivity shifts the focus to indoor tasks, highlighting the importance of accurate indoor positioning and efficient wireless power transfer for improving operational efficiency and user experience \cite{zafari2019survey}. Specifically, passive indoor positioning is particularly gaining attention because it improves work efficiency by reducing reliance on manual operations. Unlike active system, base stations (BS) can receive signals from passive targets (PT) within the environment. These signals are passively received and processed to ascertain the targets' precise location, then the system provides relevant services based on the target location, such as shopping mall guides, blind navigation, smart factories, etc. However, traditional passive positioning schemes are limited by environmental characteristics. In this paper, we proposes a RF-band resonant beam positioning system (RF-RBPS) as shown in Fig.~\ref{fig_1}, which can achieve high-precision passive positioning without the need for complex beam control and environmental estimation.

\begin{table*} [htbp]
\textcolor{black}{	
	\caption{Comparison of indoor positioning schemes}
	\renewcommand{\arraystretch}{0.95}
	\centering
	\begin{tabular}{m{2cm}<{\centering}|m{1.2cm}<{\centering}|m{9cm}<{\centering}|m{4cm}<{\centering}} \hline
		\textbf{Ref} & \textbf{Scheme} &  \textbf{Features and Accuracy}&\textbf{Application Scenario} \\
		\hline
		Le Dortz \newline et al.\cite{6288374} & WiFi & The WiFi-based AP collects signal strength and performs localization using the fingerprinting method, with an error of 2.4m. & Interactive building maps, real-time indoor tracking, etc.\\
  		\hline
		Kim \textit{et al.} .\cite{8319264} & UWB & By analyzing the signal strength in multipath, NLOS positioning is achieved with a positioning error of submeter level. & Obstacle detection, multipath signal processing, etc.\\
  		\hline
		Ji \textit{et al.} \cite{7224764} & BLE & By using BLE beacons and APs to sense targets, the number of beacons determines the accuracy, which ranges from 5 to 40m & Public safety, location-based services, mobile payments, etc.\\
  		\hline
		Ma \textit{et al.} \cite{9052482} & RFID & RFID tags emit data, and RFID readers use predefined RF and protocol to read data, achieving sub-meter accuracy. & Logistics, personal positioning, smart environments, etc.\\
  		\hline
		Bianchi \newline \textit{et al.} \cite{bianchi2018rssi} & ZigBee & Estimates distance between two or more ZigBee sensor devices based on RSSI, achieving up to 98\%  accuracy. & Ambient assisted living, user behavior analysis, etc.\\
  		\hline
		Zhu \newline \textit{et al.} \cite{zhu2023positioning} & VLC & Not relying on inertial measurement units (IMU), suitable for users at different tilt angles, can achieving sub-meter level accuracy. & Device tracking, smart Home, robot positioning, etc.\\
  		\hline
		Liu \textit{et al.} \cite{liu2023passive} & OF-RBPS & Based on the inherent features of self-alignment and energy concentration, the accuracy can reach within 1 cm within a 15$^\circ$ FoV at a distance of 2m. & Virtual reality, unmanned systems, etc.\\		
		\hline
	\end{tabular}}
	\label{tab:ref}
\end{table*}

\textcolor{black}{With the diversified development of mobile intelligent devices, various indoor positioning solutions have emerged, various indoor positioning solutions have emerged.} In the RF positioning scheme, WiFi positioning utilizes access points within the wireless network coverage area to estimate the position of objects. Bluetooth positioning is determined by measuring the strength and propagation time of Bluetooth signals. Ultra wide band (UWB) positioning is a more accurate method because it uses extremely short pulses to measure distance, providing centimeter level accuracy \cite{liu2007survey}, \cite{subedi2020survey}.
Radio frequency identification (RFID) is a wireless, non-contact technology that operates through the back-scatter communication of RFID tags, as well as the operation on RFID readers and middle-ware, thereby achieving automatic identification \cite{tesoriero2010improving}. \textcolor{black}{Antenna arrays have been proven to significantly improve positioning accuracy and resolution \cite{aliyazicioglu2008sensitivity}, which is attributed to the inverse relationship between the physical size of the antenna and the beam width. After collecting signals, the array uses array signal processing (ASP) technology to eliminate noise and interference and enhance signal quality. The spatial spectrum analysis method is commonly used in ASP, which can estimate the direction of arrival (DoA) of signals with omnidirectional distribution in space. Estimation of signal parameters via rotational invariance techniques (ESPRIT) \cite{roy1989esprit} and multiple signal classification (MUSIC) \cite{schmidt1986multiple} are two widely-used spatial spectral estimation algorithms based on the principle of eigenvalue decomposition. However, ESPRIT can only be applied to arrays with certain specific geometric shapes. In contrast the MUSIC algorithm is the most classic and widely accepted parameter estimation technique, suitable for arrays of any shape, and has extremely high resolution.
Classic RF positioning schemes can utilize pre-existing wireless networks, offering cost-effectiveness and ease of deployment \cite{alam2020device}, but faces issues such as shadowing, multipath, and omnidirectional radiation loss \cite{yuan2023three}, \cite{pandey2020adaptive}, and require a substantial number of wireless nodes to function effectively \cite{davidson2016survey}, \cite{konings2018device}.} 

\textcolor{black}{Regarding optical positioning scheme, visible light communication (VLC) uses LED lights to send and receive information to determine the position \cite{matheus2019visible}.} Infrared positioning is typically used for short distances and indoor environments, estimating positions through the reflection and reception of infrared light. Emerging optical frequency band resonant beam positioning system (OF-RBPS) is an innovative technology known for its energy concentration and self-alignment characteristics, addressing some of the limitations of traditional optical positioning methods \cite{liu2022simultaneous}. \textcolor{black}{The resonance field is formed by the superposition of in-phase waves, and waves of different phases are cancelled out through iteration. The energy exchange between two objects in the resonance state is highly efficient \cite{aoki2006observation}. Thus, the resonant beam system (RBS) is initially applied to optical wireless energy transmission \cite{zhang2018distributed}.} \textcolor{black}{The} transmitter and receiver of RBS have two reciprocal reflectors with a cat-eye-like structure, and the transmitter has a gain medium. Photons oscillate between the two reciprocal reflectors and are amplified through the gain medium.
\textcolor{black}{The authors of \cite{liu2023passive} based on this structure, obtained the coordinates of the target through geometric analysis of the light spot at the BS. However, like other optical positioning systems, RBS still has unresolved issues such as poor environmental robustness, low photoelectric conversion efficiency, and poor non-line-of-sight (NLOS) performance.}

\textcolor{black}{In TABLE~\ref{tab:ref}, we have detailed the common indoor positioning technologies, from which we can draw the following conclusions: i) Indoor positioning technologies are progressing towards higher precision; ii) RF signals are inexpensive and easy to deploy, but they are significantly affected by multipath effects and omnidirectional scattering; iii) High-frequency signals have shorter wavelengths and higher positioning resolution, but they suffer from poor penetration.}
 
\textcolor{black}{In this paper, we focus on the efficacy of the proposed RF-RBPS in indoor positioning contexts.} Specifically, we employ the retro-directive array at both ends of the BS and the PT, between which the RF resonant beam will automatically established once the threshold is satisfied. Then, we adopt the MUSIC algorithm to estimate \textcolor{black}{the} direction of arrival (DoA) based on the signal reflected from the PT \textcolor{black}{and} then received by the BS. This system does not necessitate active signal transmission from the target, thereby simplifying system complexity and reducing energy requirements. \textcolor{black}{It has the following features: i) The self-establishing link can automatically align with the target without the need for complex beam control; ii) The energy-concentrated resonant beam enhances the system's positioning accuracy.} Consequently, the design not only presents a viable alternative to existing technologies but also offers fresh perspectives for further research and development in indoor positioning.

The primary contributions of the paper are as follows.
\begin{itemize}
    \item [c1)] We propose a resonant beam system for position estimation within the radio frequency spectrum. \textcolor{black}{Compared with traditional RF positioning systems, this design eliminates the need for active signal transmission of targets and does not require complex beam control algorithms. By using a self aligned resonant beam between the base station and the target for direction estimation, millimeter level positioning accuracy can be achieved.}
    
    \item [c2)] \textcolor{black}{We constructed an analysis model of RF-RBPS using antenna principles and EM propagation theory, and used the MUSIC algorithm to estimate the DoA of the signal reflected from passive target. The results verified the feasibility and effectiveness of the proposed RF-RBPS, providing guidance for passive system performance evaluation and optimization.}
\end{itemize}

The remainder of this paper is structured as follows. In Section II, we describe the system architecture. In Section III, the analytical model of the system is introduced, mainly including the energy cycle model and the positioning model. In Section IV, we present the construction of the simulation platform and evaluate the proposed RF-RBPS through numerical analysis. \textcolor{black}{Finally, draw conclusions and prospects in Section V.}

\section{System Overview}
\textcolor{black}{In this section, we initially present the system design of the RF-RBPS. 
 Subsequently, we explained the passive localization scheme based on resonance mechanism.}
 
\subsection{System Structure of RF-RBPS}
\begin{figure*}[!t]
\centering
\includegraphics[width=6in]{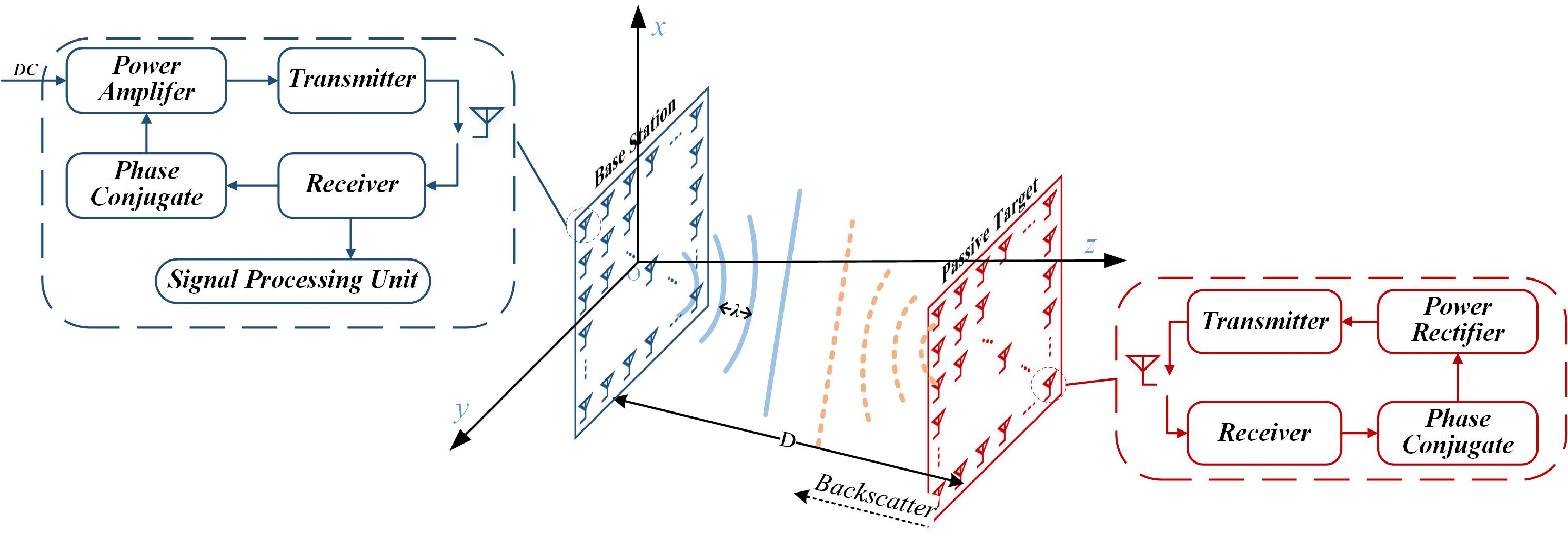}
\caption{\textcolor{black}{The structure of the RF-band resonant beam positioning system(RF-RBPS).}}
\label{rbs1}
\end{figure*}

\textcolor{black}{The RF-RBPS relies on phase conjugation circuits within retro-directive antenna to redirect incident electromagnetic (EM) waves along their original trajectories without the need of knowing the source direction \cite{leong2003moving}, \cite{dardari2023establishing}, and the retro-directive antenna is widely used in wireless power transmission \cite{wang2014wireless}, wireless communication \cite{miyamoto2002retrodirective}, and wireless sensor \cite{islam2017designing}.}
As shown in Fig.~\ref{rbs1}, the RF-RBPS consists of a base station and a passive target. The BS includes a uniform plane array (UPA) connected to a DC power source, comprised of retro-directive antenna with each element equipped with power amplifiers, transceivers, phase conjugation circuits, and signal processing units. The PT is a battery-free retro-directive antenna , only equipped with transceivers, power dividers, and phase conjugation circuits. 

In its initial state, the BS broadcasts EM waves into free space. Upon receiving the EM waves from the BS, the PT retrieves a portion of the energy for \textcolor{black}{its operation} via the power divider and returns a small portion of the energy to the BS through the retro-directive antenna. Subsequently, the BS localizes it using the signal processing unit. Then, the BS conducts phase conjugation on the received EM wave through the conjugate circuit, and amplifies the power of this EM wave to compensate for losses during transmission and the output power of the \textcolor{black}{PT. Finally, the amplified energy is \textcolor{black}{returned} to the PT, completing a power cycle.}

\subsection{Passive Positioning Principle}

\begin{figure}
    \centering
    \includegraphics[width=0.8\linewidth]{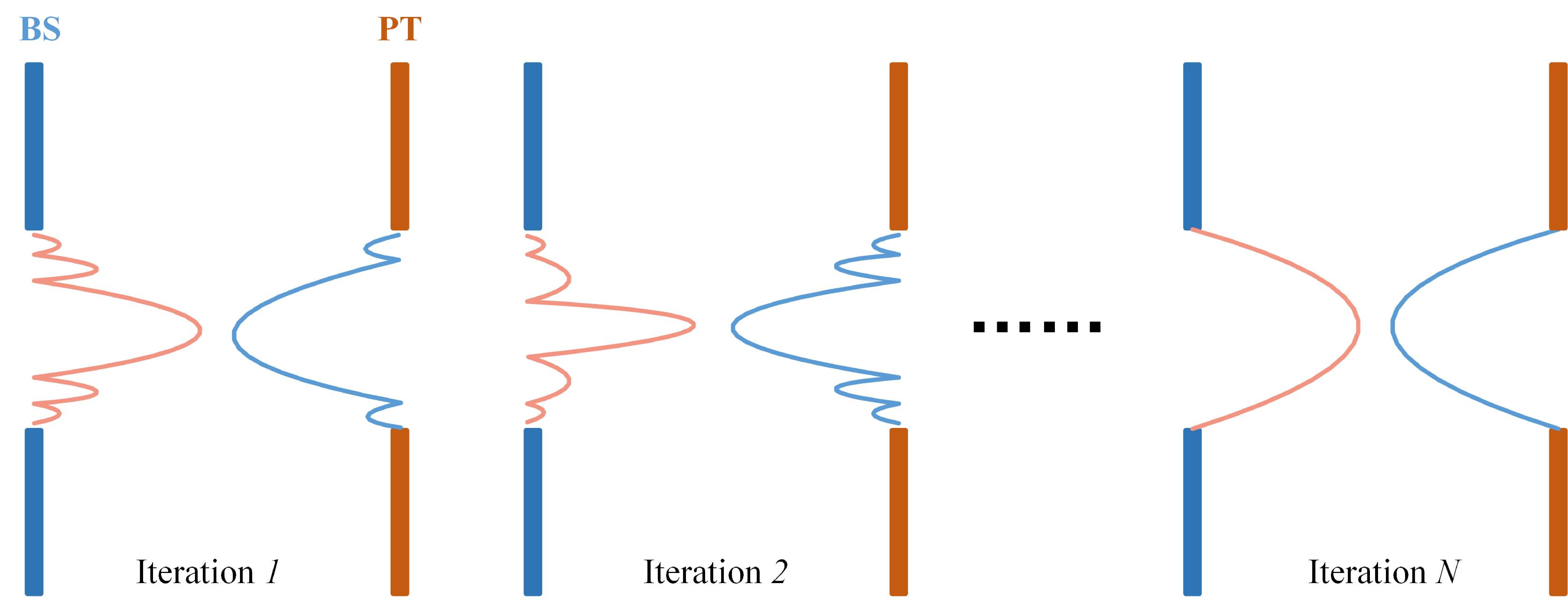}
    \caption{\textcolor{black}{The process of self-replication and reaching steady state of electromagnetic field distribution.}}
    \label{fig:rbp3}
\end{figure}

\textcolor{black}{The resonance mechanism facilitates passive positioning. 
Initially, the phases of low-power EM waves radiating omnidirectionally in space from the retro-directive antenna array of the BS is uncertain.
When part of the waves reaches the PT's retro-directive antenna array, some are reflected to the BS, re-amplified, and re-transmitted towards the PT. During the continuous oscillation of EM waves between the BS and the PT, the field distribution gradually self-replicates and reaches a steady state, as shown as Fig.~\ref{fig:rbp3}, at this point, the phase distribution of the BS also reaches stability. Although losses cannot be avoided during this process, they can be compensated for through the amplifier of the BS. Ultimately the back and forth EM waves form resonance, which can achieve energy concentration and self-alignment. This is crucial for high-precision passive positioning.}

\textcolor{black}{After the RF-RBPS reache a steady state, the BS can obtain PT angle information by receiving the phase difference of the echo reflected from the PT through each array element. Firstly, from the signals received by the BS array elements, phase differences can be obtained due to different propagation path lengths.} Then, by accurately measuring these phase differences and combining them with the array layout and signal wavelength, the phase difference information can be converted into the azimuth and elevation angles of the PT. Finally, using spatial spectrum estimation techniques such as \textcolor{black}{the} MUSIC algorithm, the spatial spectrum of the signal is constructed by analyzing phase differences, and the peak position is searched on the spectrum to accurately estimate the direction of the PT. \textcolor{black}{In traditional RF-band active positioning systems (RF-APS), the target generally needs to actively emit signals, which} suffer from low directivity, resulting in weaker signal strength, lower signal-to-noise ratio \textcolor{black}{(SNR)} and subsequent poor positioning accuracy.

\section{ANALYTICAL MODEL}
In this section, leveraging EM principles and antenna theory, we formulate an RF-RBPS power-cycle model. Considering the orthogonal relationship between the signal power and the noise power received at the BS, we employ the MUSIC algorithm to estimate the DoA of the PT based on the established RF-RBPS power model. Finally, we determine the estimated position of the target and assess the accuracy using the root mean square error (RMSE).

\subsection{Power Cycling Model}

\subsubsection{Electromagnetic Field Description}
In the EM field, power density represents the power received per unit area at a specific location, and is a crucial parameter for evaluating the efficiency and performance of wireless energy transmission \cite{jackson1999classical}. \textcolor{black}{For the far-field region of the transmitter, the power density $W$ is typically expressed as}
\begin{equation}
\label{eq1}
W=\frac{P^{out} G(\theta,\phi)}{4\pi D^2},
\end{equation}
where $P^{out}$ is the average power radiated by the antenna, and $G(\theta,\phi)$ is the antenna gain in the direction of the point being considered, $\theta$ is the elevation angle, $\phi$ is the azimuth angle, and $D$ is the distance between the BS and PT.

We can use Poynting vector to represent the cross-coupling relationship between electric and magnetic fields in Maxwell's equations
\begin{equation}
\label{eq2}
    \vec{S}=\vec{E}\times\vec{H},
\end{equation}
\textcolor{black}{where $\vec{S}$ is the Poynting vector, $\vec{E}={\vec{E}}_{max}\cos{\omega t}$, $ \vec{H}={\vec{H}}_{max}\sin{\omega t}$, ${\vec{E}}_{max}$ and ${\vec{H}}_{max}$ are the maximum values of the electric and magnetic, $\omega$ is the angular frequency of the EM waves.}

Poynting vector can denote the direction of the EM wave propagation, also known as power flow density and the magnitude of the Poynting vector is equal to the power density $W$. According to the energy transfer theorem, the power density can be expressed as
\begin{equation}
    \label{eq3}
W=\langle \vec{S}\rangle =\frac{|\vec{E}_{max} |^2}{2\mu_0 c}=\frac {\epsilon_0 c}{2}|\vec{E}_{max} |^2,
\end{equation}
where $c$ is the speed of the EM wave, $\mu_0$ is the magnetic permeability, and $\epsilon_0$ is the electrical permittivity \cite{ulaby2013fundamentals}. 

\subsubsection{Electromagnetic Waves Transmission Model}
We assume that the BS has $M$ antennas and the PT has $N$ antennas. According to (1) and (3), we can express the transmission electrical field of the $m$-th antenna in BS as
\begin{equation}
   E_{\mathrm{BS},m}^{out} = \sqrt{\frac{{\mu_0 c P_{{\mathrm{BS},m}}^{out} G_{{\mathrm{BS},m}}}}{{2\pi \textcolor{black}{D_{m,n}}^2}}}e^{i\left(kD_{m,n}+\alpha_0\right)},
\end{equation}
where $P_{{\mathrm{BS},m}}^{out}$ is the average power radiated by the $m$-th antenna in BS, $G_{{\mathrm{BS},m}}$ is the $m$-th antenna gain in BS, $k=\frac{2\pi}{\lambda}$ is the wave number, $\lambda$ is the wavelength, $\alpha_0$ is the initial phase, and $D_{m,n}$ is the distance between the $m$-th antenna in BS and the $n$-th antenna in PT.

When multiple antennas simultaneously radiate EM waves with the same frequency, these waves can be superimposed using the principle of vector addition. Therefore, the electric field of the BS with $M$ antennas at the $n$-th antenna of the target can be expressed as
\begin{equation}
    E_\mathrm{BS}^{out}=\sum_{m=1}^{M}{\sqrt{\frac{{\mu_0cP}_{{\mathrm{BS},m}}^ {out}G_{{{\mathrm{BS},m}}}}{2\pi D_{m,n}^2}}e^{i\left(kD_{m,n}+\alpha_0\right)}}.
\end{equation}

\subsubsection{Electromagnetic Waves Receiving Model}
The power of the EM waves received by the $n$-th antenna of the PT from the BS is determined by the power density and the effective aperture of the receiving \textcolor{black}{antenna, which} is
\begin{equation}
{P_{\mathrm{PT},n}^{in}=W_\mathrm{BS}^{out}A}_\mathrm{eff},
\label{eq6}
\end{equation}
where $W_{\mathrm{BS}}^{out}$ is the power density of the EM waves transmitted by BS at the $n$-th antenna of the \textcolor{black}{PT, $A_\mathrm{eff}$} is the effective aperture of the receiving antenna \cite{balanis2016antenna}, and
\begin{equation}
    A_\mathrm{eff}=\frac{G_\mathrm{PT}\lambda^2}{4\pi},
\label{eq7}
\end{equation}
where $G_\mathrm{PT}$ is the antennas gain of the PT. Substituting the formula to (6), and combine (3) we can get
\begin{equation}
P_{\mathrm{PT},n}^{in}=\frac{G_\mathrm{PT}\lambda^2\textcolor{black}{{E_{{\mathrm{BS},n}}^{out}}}^2}{8\pi\mu_0c},
\end{equation}
where \textcolor{black}{$E_{{\mathrm{BS},n}}^{out}$} is electric field intensity from the BS to the antenna. The received phase of the antenna can be expressed by
\begin{equation}
\alpha_{\mathrm{PT},n}=\arg(E_{{\mathrm{BS},n}}^{out}).
\end{equation}

The receiving power of each antenna in the target array can be expressed by the superposition of field as
\begin{equation}
    P_{\mathrm{PT},n}^{in}=\frac{\lambda^2}{16\pi^2}\left|\sum_{m=1}^{M}{\sqrt{\frac{P_{{\mathrm{BS},m}}^{out}G_{{\mathrm{BS},m}}G_{\mathrm{PT},n}}{D_{m,n}^2}}e^{i\left(kD_{m,n}+\alpha_m\right)}}\right|^2.
\end{equation}

Furthermore, the total receiving power of the PT can be expressed by 
\begin{equation}
\label{eq12}
    P_\mathrm{PT}^{in}=\frac{\lambda^2}{16\pi^2}\sum_{n=1}^{N}\left|\sum_{m=1}^{M}{\sqrt{\frac{P_{{\mathrm{BS},m}}^{out}G_{{\mathrm{BS},m}}G_{\mathrm{PT},n}}{D_{m,n}^2}}e^{i\left(kD_{m,n}+\alpha_m\right)}}\right|^2.
\end{equation}

\begin{figure}[!t]
\centering
\includegraphics[width=3in]{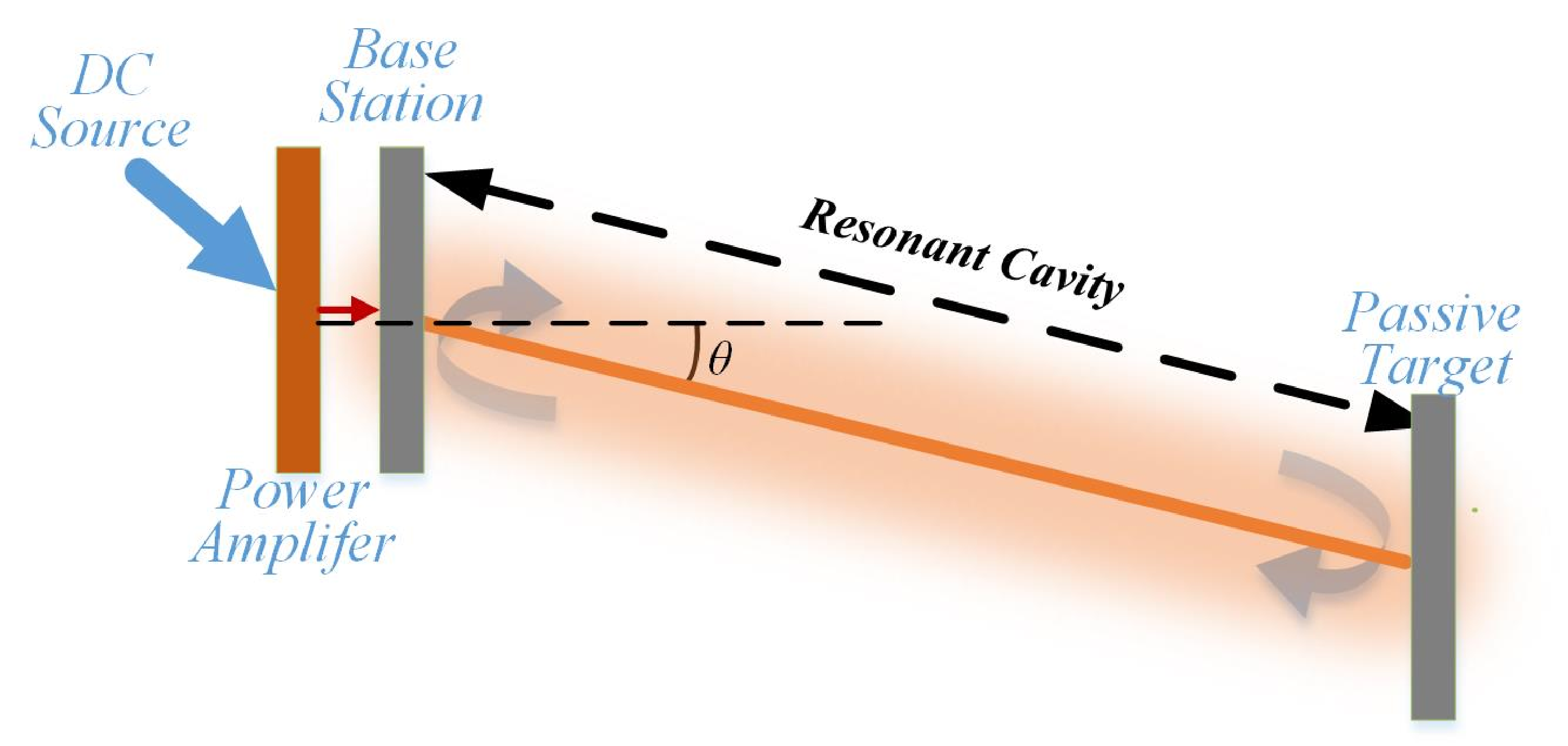}
\caption{Power cycle structure based on resonance principle.}
\label{rbs2}
\end{figure}

\subsubsection{Power Cycle Process}
In this subsection, we provide a detailed introduction to the principle of RF-RBPS. As shown in Fig.~\ref{rbs2}, EM waves are not unidirectional in the RF-RBPS\textcolor{black}{, the} cycle from the BS to the PT and then reflected back by the PT to the BS is called an iteration. The BS is connected to a DC power source and amplifies the input EM waves through a power amplifier, it can be expressed as,
\begin{equation}
P_\mathrm{BS}^{out}=f_\mathrm{PA}(P_\mathrm{BS}^{in}),
\end{equation}
where the $f_\mathrm{PA}(.)$ is the amplification function, and the amplification factor of the amplifier is solely dictated by the input power, obviating the need for auxiliary control mechanisms. At lower input powers, the amplification factor is maintained constant. Conversely, as the input power increases, the amplification factor progressively diminishes, constrained by the maximal output power threshold. Upon receiving the EM waves from the BS, the phase of the EM waves received by the PT can be represented as,
\begin{equation}
    \alpha_\mathrm{RDA}^{in}=kD+\alpha_0.
\end{equation}

Note that there are all possibilities for the distance $D$ from the receiving antenna to the transmitting antenna. \textcolor{black}{Through} the phase conjugate by the phase conjugate circuit in the retro-directive array, the input phase $\alpha_\mathrm{RDA}^{in}$ becomes the output phase
\begin{align}
        \alpha_\mathrm{RDA}^{out}&=-\alpha_\mathrm{RDA}^{in}+\alpha'\nonumber \\
        &=-kD-\alpha_0+\alpha',
\end{align}
where $\alpha'$ is the phase delay of the conjugate circuit\cite{khang2017microwave}.
The \textcolor{black}{retro-directive antenna} can eliminate the phase differences in the near field through the method of phase conjugation.
In RF-RBPS, the PT can reflect the received power in a \textcolor{black}{particular power feedback ratio} $\gamma$, 
\begin{equation}
    P_\mathrm{PT}^{out}=\gamma P_\mathrm{PT}^{in}.
\end{equation}

In order to ensure the efficient operation of the power division system, the power feedback ratio $\gamma$ must be properly chosen. Specifically, a lower $\gamma$ results in insufficient power return to the transmitter, causing the system to operate at a low transmission power. Conversely, a higher $\gamma$ forces all the transmitting antennas to operate at peak power, which negates the benefits of power adjustment and limits efficiency. The optimization of $\gamma$ should be subject to specific constraints to ensure the balance between power modulation and system efficiency. This balance can be formally expressed as
\begin{equation}
    \textcolor{black}{G_T \gamma \geq \zeta_T.}
    \label{eq16}
\end{equation}

\textcolor{black}{In (16), $G_T$ represents the total amplification gain of all antennas in the BS at the $T$-th iteration, and $\zeta_T$ represents the total round trip path-loss factor at the $T$-th iteration, which is determined by the round-trip transmission efficiency between the BS and the PT, i.e. $\zeta=1/(\eta \eta_{\mathrm{pt}})$. When $T \to \infty$, RF-RBPS has reached a stable operating state, and the gain and loss of the system are offset. At this point, the efficiency of the system's power transmission from BS to PT can be expressed as,}
\begin{align}
    \eta = \frac{P_\mathrm{PT}^{in}}{P_\mathrm{BS}^{out}}
    =\frac{\frac{\lambda^2}{16\pi^2} \left| \sum\limits_{m=1}^{M} \sum\limits_{n=1}^{N} \frac{\sqrt{G_{{\mathrm{BS},m}} G_{\mathrm{PT},n} P_{{\mathrm{BS},m}}^{out}}}{D_{m,n}} e^{j(kD_{m,n} + \alpha_m)} \right|^2 }{\sum\limits_{m=1}^{M} P_{{\mathrm{BS},m}}^{out}},
\end{align}
where $P_\mathrm{PT}^{in}$ represents the input power at the PT with $T$ iterations, and $P_\mathrm{BS}^{out}$ represents the output power from the BS at iteration $T$.

\subsection{DoA Estimation with MUSIC Algorithm}

\begin{figure}[!t]
\centering
\includegraphics[width=2.5in]{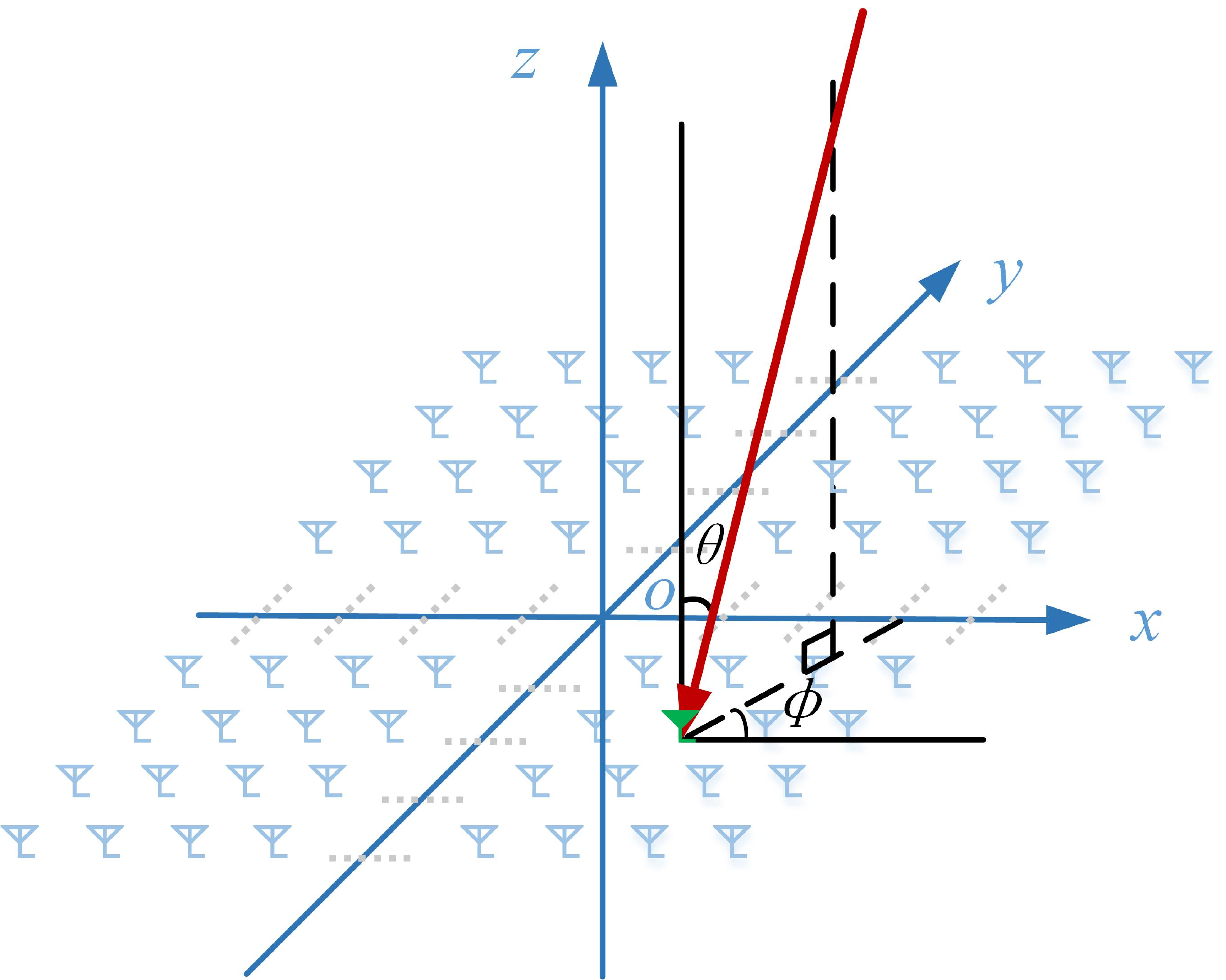}
\caption{Schematic diagram of elevation angle and azimuth angle of UPA.}
\label{DoA}
\end{figure}

\textcolor{black}{Since the resonant beam reflected from the PT is received by the retro-directive antenna array in BS, the EM waves propagating in a 3-D direction will be sampled many discrete angles. As depicted in Fig.~\ref{DoA}, these discrete angles correspond to the elevation angle $\theta$ and azimuth angle $\phi$ at a macro level. Subsequently, it is assumed that both the source signal and noise signal arrive from each minor direction. Using the MUSIC algorithm, we can determine the $\theta$ and $\phi$, as the DoA of the maximum power signal}.

\textcolor{black}{As the BS consists of a UPA with $M$ elements, the signal received by can be expressed as}
\begin{equation}
\textcolor{black}{\mathbf{Z}=\mathbf{A} \mathbf{P_S} + \mathbf{P_N}},
\end{equation}
\textcolor{black}{where $\mathbf{P_N} = [P_{\mathrm{bs},1}^{n}, P_{\mathrm{bs},2}^{n}, \ldots, P_{\mathrm{bs},M}^{n}]^T$ is the noise power at each antenna of the BS, and $\mathbf{P_S} = [P_{{\mathrm{bs},1}}^{in}, P_{{\mathrm{bs},2}}^{in}, \ldots, P_{{\mathrm{bs},M}}^{in}]^T$ is the signal power received by the BS from the PT}. Here, $\mathbf{A} = [\alpha_1(\theta, \phi), \alpha_2(\theta, \phi), \ldots, \alpha_M(\theta, \phi)]$ is the array manifold matrix, where the
\textcolor{black}{$\alpha_j$} is the directional vector of the target transmission signal received by the $j$-th element of the BS, and can be denoted as
\begin{equation}
    \alpha_j(\theta,\phi) =[e^{i\omega\tau_{1j}(\theta,\phi)}, e^{i\omega\tau_{2j}(\theta,\phi)}, \ldots,e^{i\omega\tau_{Mj}(\theta,\phi)}],
\end{equation}
where $\omega=\frac{2\pi{c}}{\lambda}$, $\tau_m$ is the delay of the $m$-th element relative to the direction of wave arrival, defined by elevation angle $\theta$ and azimuth angle $\phi$ as shown in Fig.~\ref{DoA}, i.e.,
\begin{equation}
    \tau_m(\theta, \phi)=\frac{1}{c}(x_m\sin{\theta}\cos{\phi}+y_m\sin{\theta}\sin{\phi}+\textcolor{black}{z_m\cos{\theta}}),
\end{equation}
the covariance matrix $\mathbf{R}\in\mathbb{C}^{M\times M}$ of BS receiving signal is
\textcolor{black}{
\begin{align}
    \mathbf{R}_Z&=\mathbb{E}[\mathbf{ZZ}^H]\nonumber \\
    &=\mathbf{A}\mathbb{E}[\mathbf{P_SP_S}^H]\mathbf{A}^H+\mathbf{A}\mathbb{E}[\mathbf{P_SP_N}^H]+\mathbb{E}[\mathbf{P_NP_S}^H]\mathbf{A}^H\nonumber \\
    &+\mathbb{E}[\mathbf{P_NP_N}^H].
\end{align}}
where $\mathbb{E}$ denotes the expectation operator, $H$ is the conjugate transpose, and the noise is independent with the signal, and can be considered as additive gaussian white noise (AGWN) with a noise power of $\sigma_N^2$. Therefore, we can obtain $\mathbb{E}[{\mathbf{P_SP_N}}^H]=\mathbb{E}[\mathbf{P_NP_S}^H]=0$, $\mathbb{E}[\mathbf{P_NP_N}^H]=\sigma_N^2\mathbf{I}_M$, where $\mathbf{I}_M$ is an $M$-dimensional identity matrix, and $\mathbf{R}_P=\mathbb{E}[{\mathbf{P_SP_S}}^H]$. Thus, the above covariance matrix can be derived as
\begin{equation}
   \mathbf{R}_Z=\mathbf{A}{\mathbf{R}}_P\mathbf{A}^H+\sigma_N^2\mathbf{I}_M.
\end{equation}
where $\mathbf{R}_P$  is a full rank matrix with rank $K$, and $\mathbf{A}{\mathbf{R}}_P\mathbf{A}^H$ also is a full rank matrix with rank $K$ which is the number of PT signal sources. $\lambda_i$ is the eigenvalues.

Furthermore, the eigenvalues of matrices $\mathbf{R}_Z$ can be expressed by
\begin{align}
    \gamma_i&=\lambda_i+\sigma_N^2, \left\{i=1,2,...,K\right\},\nonumber\\
    \gamma_i&=\sigma_N^2, \left\{j=K+1,K+2,...,M\right\},
\end{align}
and the eigenvectors of matrices $\mathbf{R}_Z$ is $e_j, j=1,2,...,M$. The first $K$ elements  of $\mathbf{R}_Z$ correspond to the largest $K$ eigenvalues, the remaining elements correspond to the smallest $M-K$ eigenvalues.

We construct a noise subspace as
\begin{equation}
    V_N=[e_{K+1}, e_{K+2},...,e_{M}].
\end{equation}

The MUSIC spatial spectrum is
\begin{equation}
    \textcolor{black}{P_\mathrm{MUSIC}}(\theta, \phi)=\frac{1}{\alpha^H(\theta, \phi)V_NV_N^H\alpha(\theta, \phi)},
\end{equation}
where $\alpha$ is the directional vector, and orthogonal to $M-K$ smallest eigenvectors, i.e., signal subspace is orthogonal to the noise subspace.

In the presence of various noise interferences affecting the received signal, errors in angle estimation are inevitable. We employ the RMSE of the measured angles versus the actual angles to evaluate the azimuthal accuracy of the \textcolor{black}{target}, which is defined as:
\begin{align}
\Delta x_i &= D \sin \hat{\theta}_i \cos \hat{\phi}_i - D \sin \theta_0 \cos \phi_0, \\
\Delta y_i &= D \sin \hat{\theta}_i \sin \hat{\phi}_i - D \sin \theta_0 \sin \phi_0, \\
\Delta z_i &= D \cos \hat{\theta}_i - D \cos \theta_0, \\
\text{RMSE} &= \sqrt{\frac{1}{n} \sum_{i=1}^{n} (\Delta x_i^2 + \Delta y_i^2 + \Delta z_i^2)},
\label{eq29}
\end{align}
where $\hat{\theta}_j$ and $\hat{\phi}_j$ is the estimated elevation angle and azimuth angle obtained from the $j$-th Monte Carlo experiment, $\theta_0$ and $\phi_0$ is the precise elevation angle and azimuth angle, $n$ is the number of Monte Carlo experiments or data points.

\section{PERFORMANCE EVALUATION}
In this section, we assess PT positioning performance of the RF-RBPS. Firstly, we detail the simulation parameters and device selections. Then, we perform the comparisons upon the power concertration between the proposed RF-RBPS and the traditional RF-APS under identical conditions. We also analyze the impact of the number of iterations and the misalignment of the transceivers on the total transmission efficiency. Lastly, we evaluate the performance of the DoA estimation by assessing the RMSE of the positioning accuracy.

\subsection{Parameter Settings and Devices Selection}
\begin{table}[h]
  \textcolor{black}{
    \centering
    \caption{Parameter Setting}
    \begin{tabular}{m{3cm}<{\centering} m{2cm}<{\centering} m{2cm}<{\centering}}
        \toprule 
        \textbf{Parameter} & \textbf{Symbol} & \textbf{Value} \\
        \midrule
        Frequency & $f$ & 30GHz \\
        Wavelength & $\lambda$ & 1 cm \\
        Antenna element spacing & $\lambda/4$ & 0.25cm \\
        Antenna gain \cite{balanis2016antenna} & $G(\theta,\phi)$ & $\leq$4.97dBi \\     
        Distance from BS to PT & $d$ & (2m, 2.5m) \\
        Feedback ratio & $\gamma$ & 0.004 \\
        Array side number of BS and PT & $N_s$ & (30, 35, 40) \\
        Amplifier gain \cite{devices2023}& $G_\mathrm{PA}$ & $\leq$24dB \\
        Iterations & $T$ & 200 \\
        Monte Carlo number & $n$ & 100 \\
        \bottomrule
    \end{tabular}
    \label{tab:parameter_setting}}
\end{table}

\begin{figure}
  \centering
\subfigure[]{
	\includegraphics[width=0.45\linewidth]{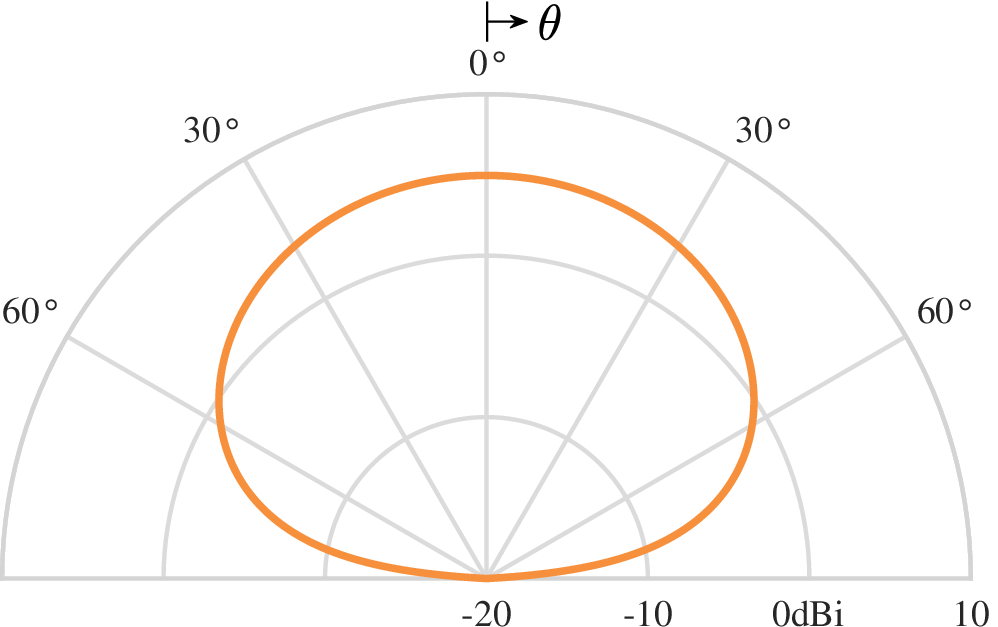}}
        \label{gain1}
\subfigure[]{
        \includegraphics[width=0.45\linewidth]{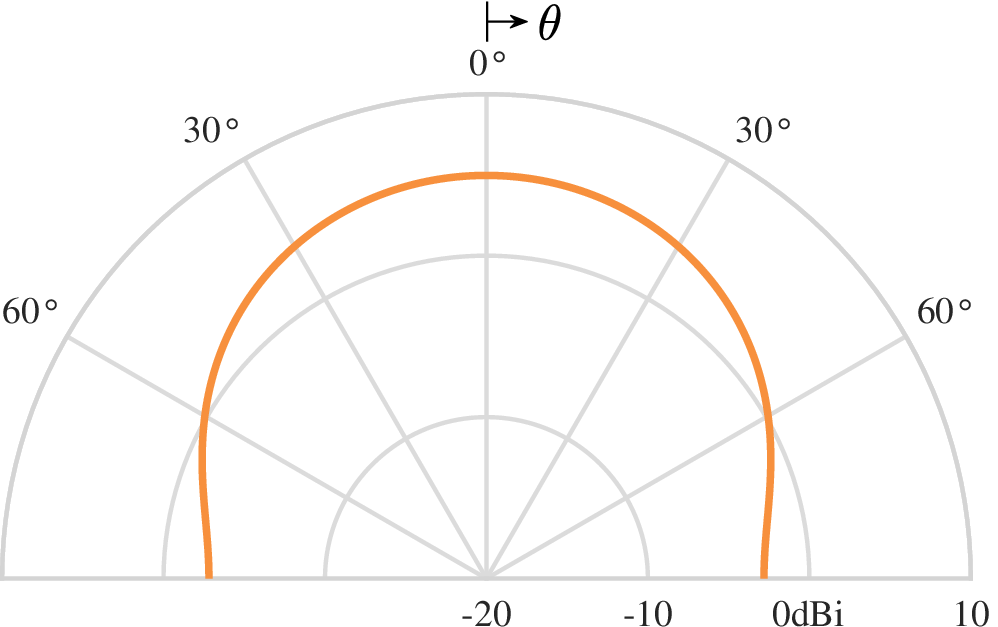}}
	\label{gain2}
  \caption{\textcolor{black}{The antenna gain $G(\theta,\phi)$. (a) $\phi=0^\circ/360^\circ$. (b) $\phi=90^\circ/270^\circ$.}}
  \label{gain}
\end{figure}

\textcolor{black}{The parameters in the simulation setting are listed in TABLE~\ref{tab:parameter_setting}. We assume that in both the RF-RBPS and the referenced RF-APS, all components including the microstrip antennas, RF power amplifiers, filters, and phase conjugation circuits should be tuned to operate at a frequency of $30\mathrm{GHz}$, corresponding to a wavelength of approximately $1\mathrm{cm}$. The sizes of the transmitting and receiving arrays are the same. The spacing between the antenna elements is $\lambda/4$, and the closest simulated distance of $2\mathrm{m}$ satisfies the far-field assumption. For the antenna gain $G_(\theta,\phi)$ of the antenna, we referred to the radiation pattern proposed in \cite{balanis2016antenna}. As shown in Fig.~\ref{gain}, which shows that the antenna has the highest gain in the main beam direction (i.e. $\theta=0$), and the maximum antenna gain is $4.97\mathrm{dBi}$.}
\begin{figure}[!t]
\centering
\includegraphics[width=0.8\linewidth]{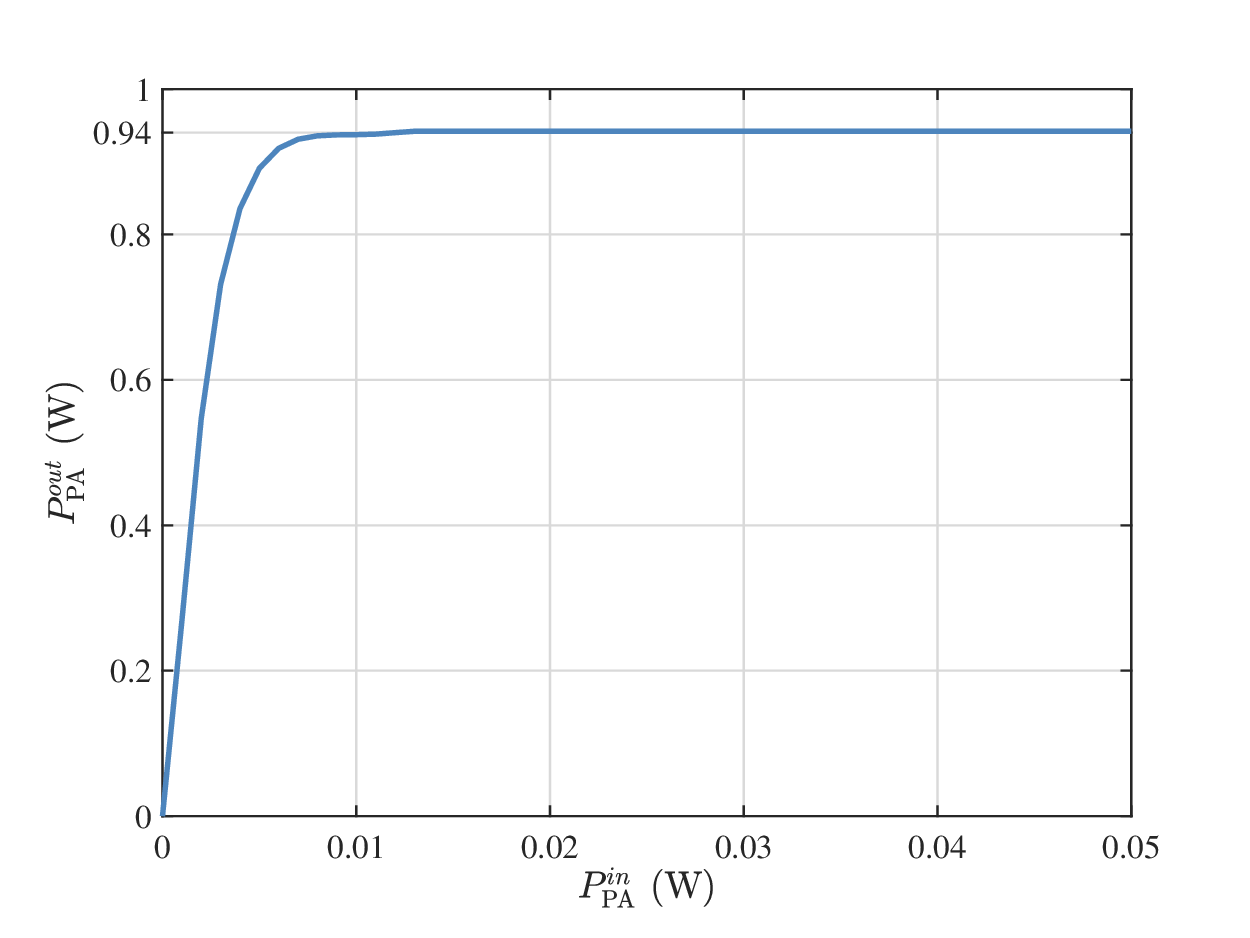}
\caption{\textcolor{black}{Output power versus input power of the BS's power amplifier.}}
\label{rfpa}
\end{figure}

\begin{figure}[!t]
\centering
\includegraphics[width=0.8\linewidth]{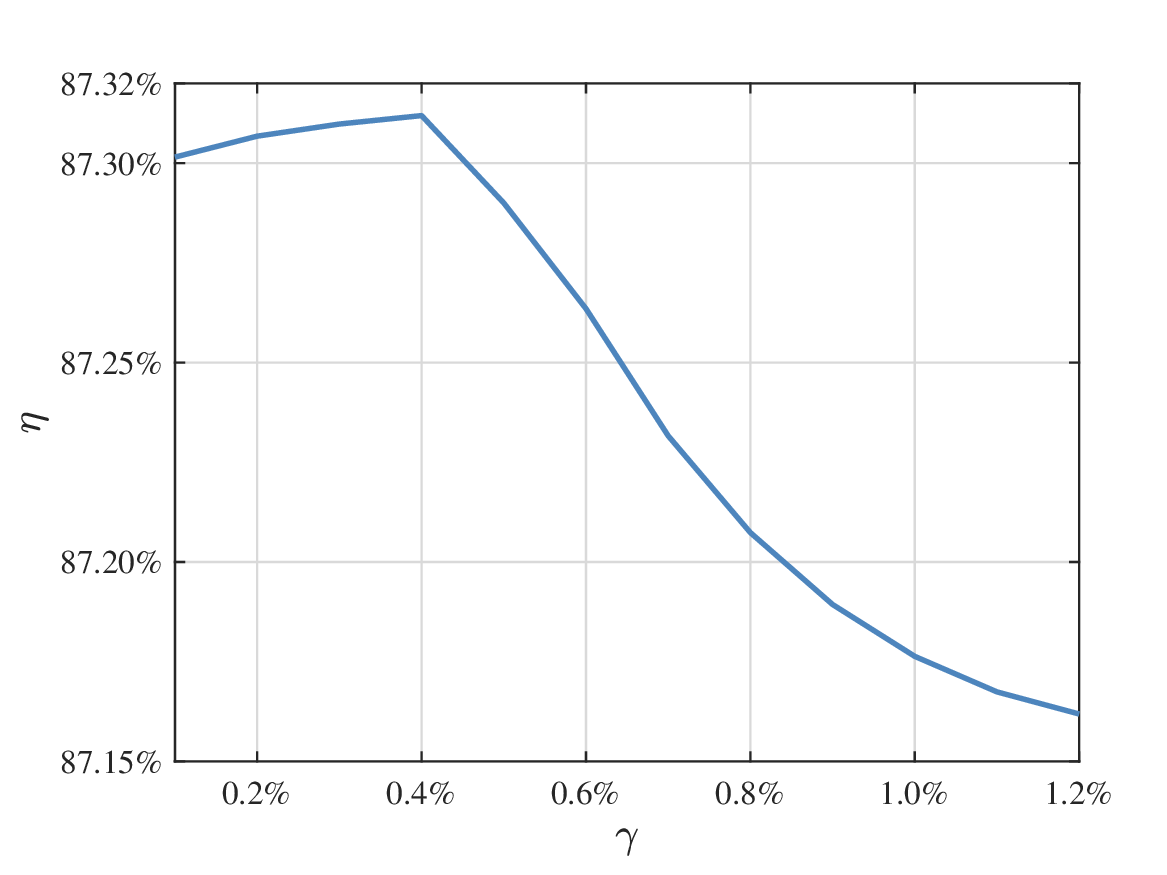}
\caption{\textcolor{black}{The impact of the power feedback ratio $\gamma$ on transmission efficiency $\eta$, where the BS and PT face-to-face and the distance $D=2\mathrm{m}$, the array size is $40\times40$.}}
\label{best_efficiency}
\end{figure}

\textcolor{black}{According to the ADI chip manual \cite{devices2023}, RF power amplifiers can operate at $30\mathrm{GHz}$.} Within its standard operating range, the output power is solely determined by the input power.
\textcolor{black}{As shown in Fig.~\ref{rfpa}, the amplification rate continues to increase with the increase of input power and ultimately remains below 1W.}

In determining the feedback ratio $\gamma$ for the power divider, we adhere to the criteria outlined in~(\ref{eq16}). Considering the aforementioned characteristics of the power amplifier, we simulated the steady-state performance for different feedback ratios with the BS and PT aligned face-to-face at a distance of $2\mathrm{m}$,  with both ends featuring with a size of \textcolor{black}{$N_s=40$}. \textcolor{black}{As shown in Fig.~\ref{best_efficiency}, with increasing $\gamma$, the transmission efficiency gradually improves, reaching its peak at $\gamma=0.004$ before starting to decrease. Therefore, we set the feedback ratio of the power divider to $\gamma=0.004$.}
\subsection{Analysis of Radiated Power}

\begin{figure*}[!t]
	\centering
 \subfigure[]{
	\includegraphics[width=0.235\linewidth]{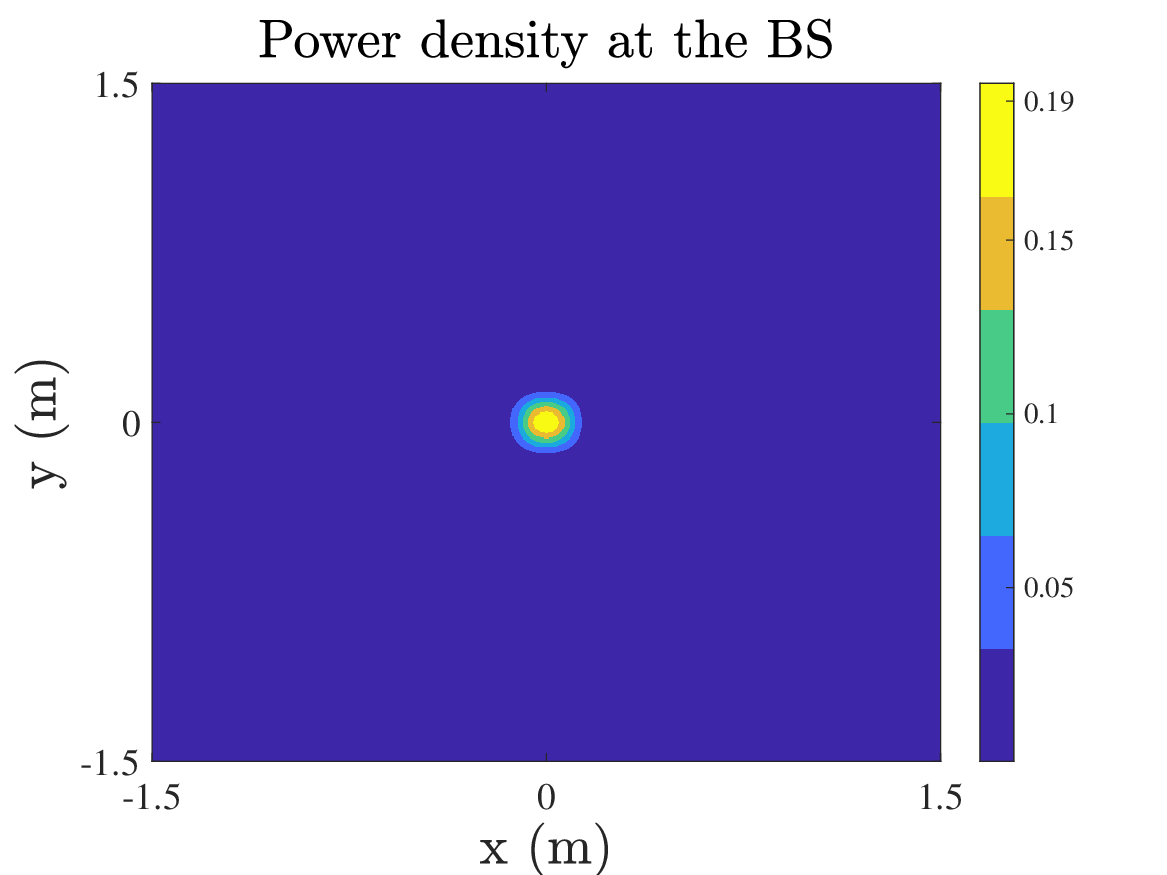}}
 \subfigure[]{
        \includegraphics[width=0.235\linewidth]{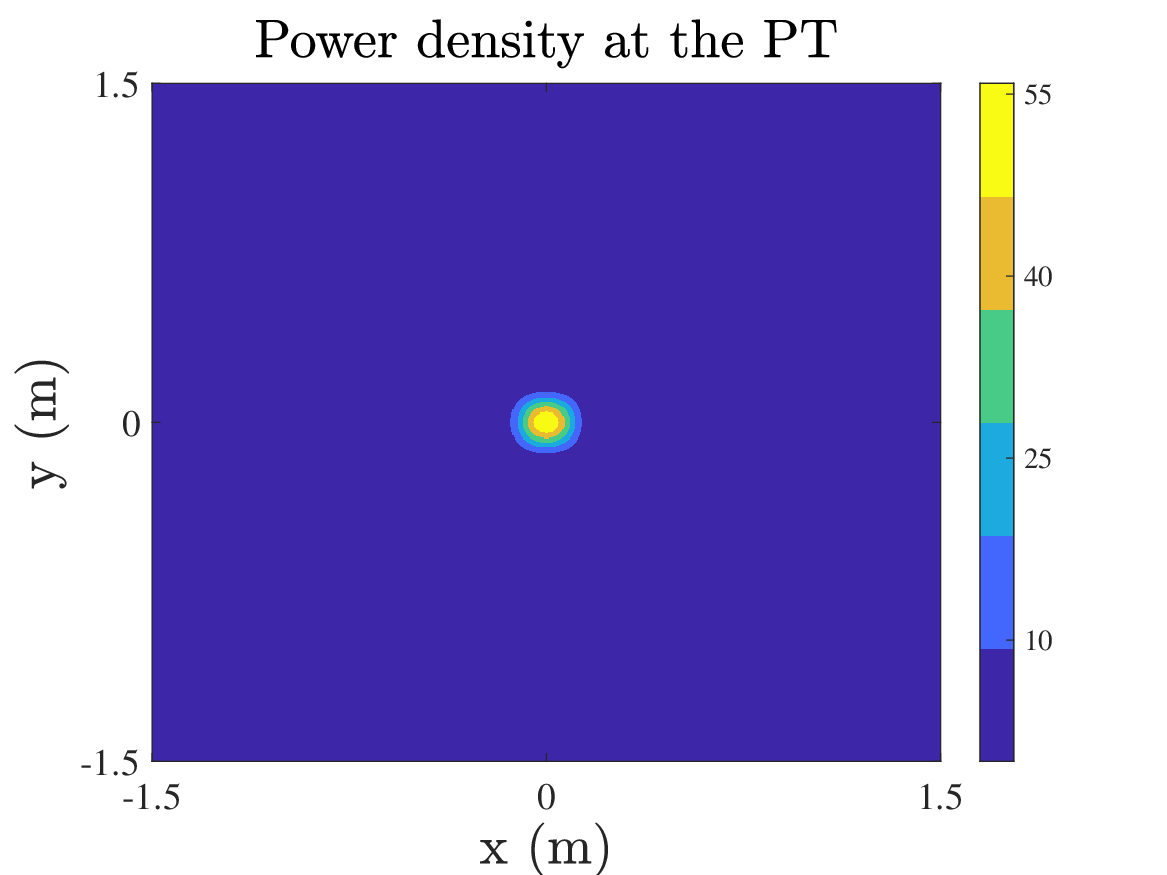}}
  \subfigure[]{
        \includegraphics[width=0.235\linewidth]{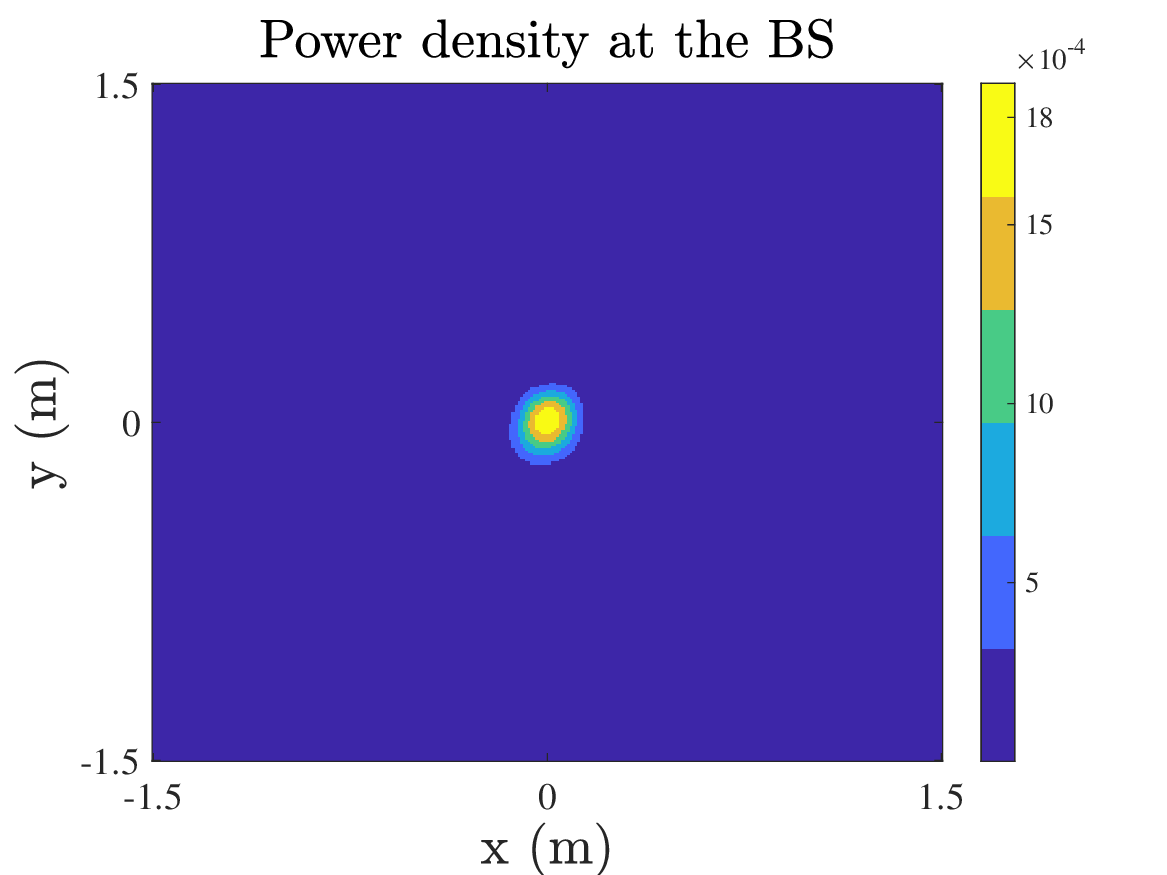}}
  \subfigure[]{
        \includegraphics[width=0.235\linewidth]{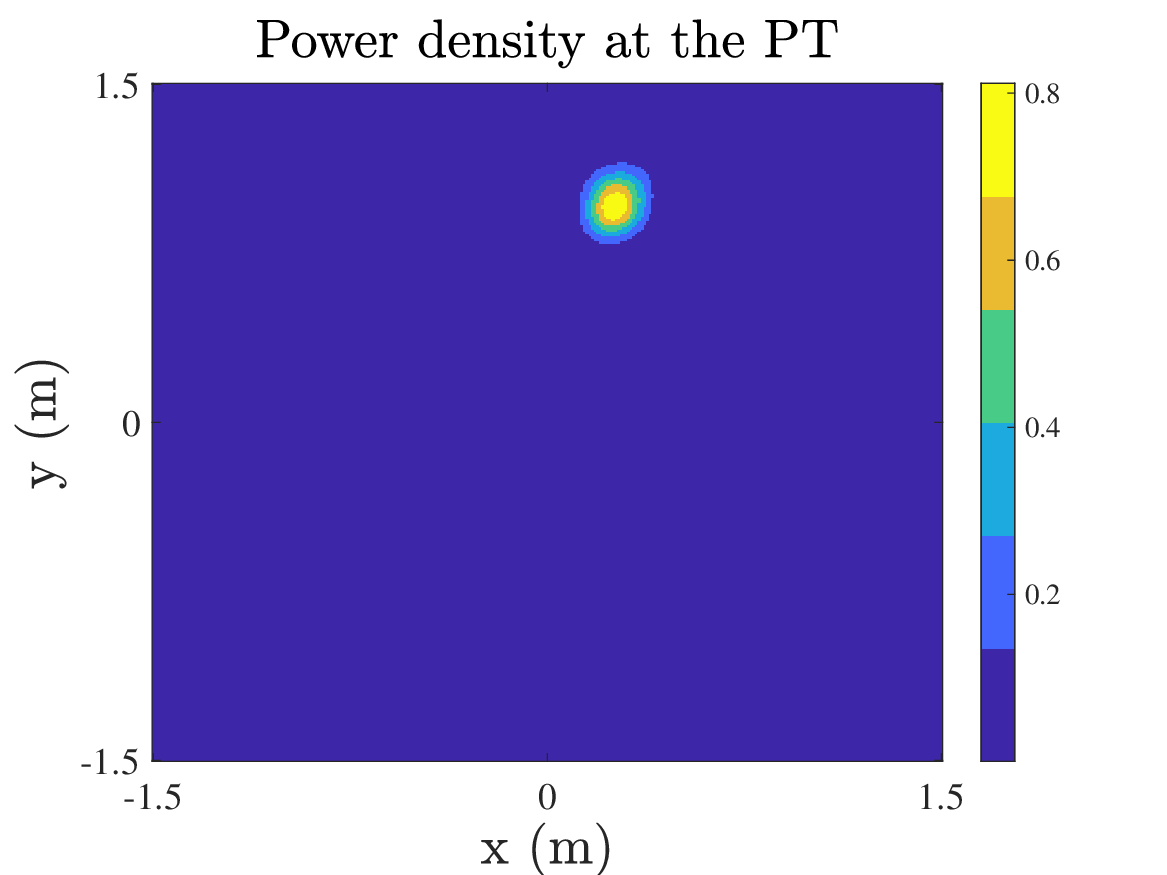}}
  \caption{\textcolor{black}{The power density at the BS and PT after EM waves reach resonance steady state ($\mathrm{W/m^2}$). (a) and (b) are the incident EM wave at the BS with $\theta=0^\circ$ and $\phi=0^\circ$ at the distance of $D=2\mathrm{m}$. (c) and (d) are the incident EM wave at the BS with $\theta=30^\circ$ and $\phi=15^\circ$ at the distance of $D=2\mathrm{m}$.}}
  \label{density}
         \label{density}
\end{figure*}

In this subsection, We analyze the influence of iteration counts on the radiated power of the PT. Lastly, we evaluate the effects of different elevation angles on system transmission efficiency.

In Fig.~\ref{density}, we simulated a scenario where both the BS and the PT array consist of $40\times40$ elements, with a distance of \textcolor{black}{$2\mathrm{m}$}. \textcolor{black}{At the outset, the BS at the origin $(0, 0, 0)$ and transmits $1\mathrm{mW}$ of power. After $200$ iterations, the system attains equilibrium.} At this juncture, for the direct alignment of the BS and the PT, the EM wave power density at the BS and at the PT are depicted in Fig.~\ref{density}(a) and Fig.~\ref{density}(b), respectively. Additionally, for an incoming EM wave at the BS with the elevation angle of $\theta=30^\circ$ and the azimuth angle of $\phi=15^\circ$, the corresponding EM wave power densities at the BS and at the PT location are presented in Fig.~\ref{density}(c) and Fig.~\ref{density}(d), respectively.  
This figure validates the proposed RF-RBPS's characteristics of energy concentration and self-alignment, with uniform electromagnetic wave reception and fewer side-lobes at both ends, which significantly improves system resolution and reduces interference from undesired directions.

\begin{figure}
    \centering
    \includegraphics[width=0.8\linewidth]{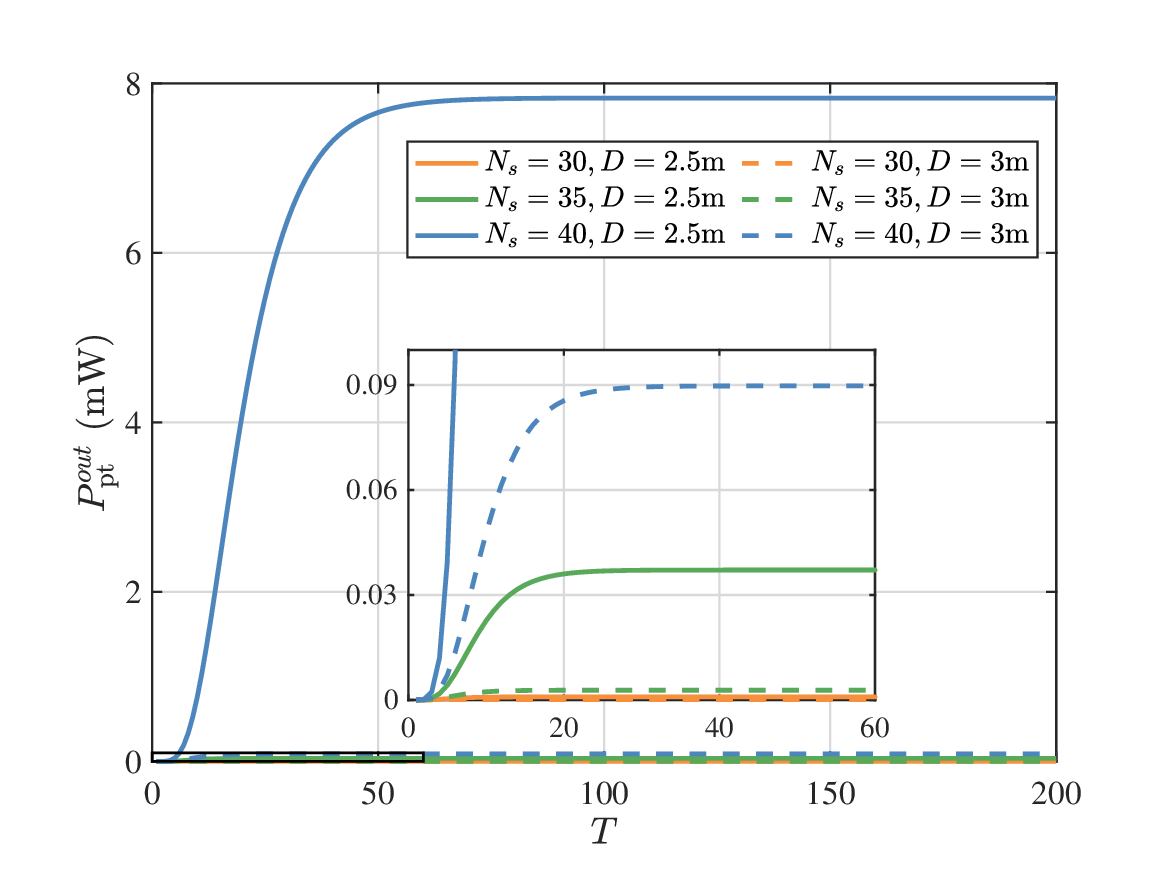}
    \caption{\textcolor{black}{The variation of PT output power $P_\mathrm{pt}^{out}$ with the number of iterations and the variation with radial distances.}}
    \label{ptout}
\end{figure}

Fig.~\ref{ptout} simulates the variation of the radiated power of the PT with the number of iterations. It can be observed that the target's radiated power starts from a lower value and steadily increases with iterations until stabilization. This is because, in RF-RBPS, the BS initially sends EM waves to the PT. After receiving the waves, the PT returns the EM waves to BS through the conjugate phase circuit for gain amplification. During multiple iterations, the system continuously adjusts the power and phase of the transmitted and received EM waves, achieving maximized transmission efficiency and balanced PT radiated power. This indicates that as the number of iterations increases, the signal strength received by the BS from the PT also continuously enhances, contributing to more accurate position estimation. Additionally, it is noted that due to limitations in array size and transmission distance, the convergence speed of power for smaller arrays or longer distances decreases, resulting in lower radiated power.

\begin{figure}
  \centering
	\includegraphics[width=0.8\linewidth]{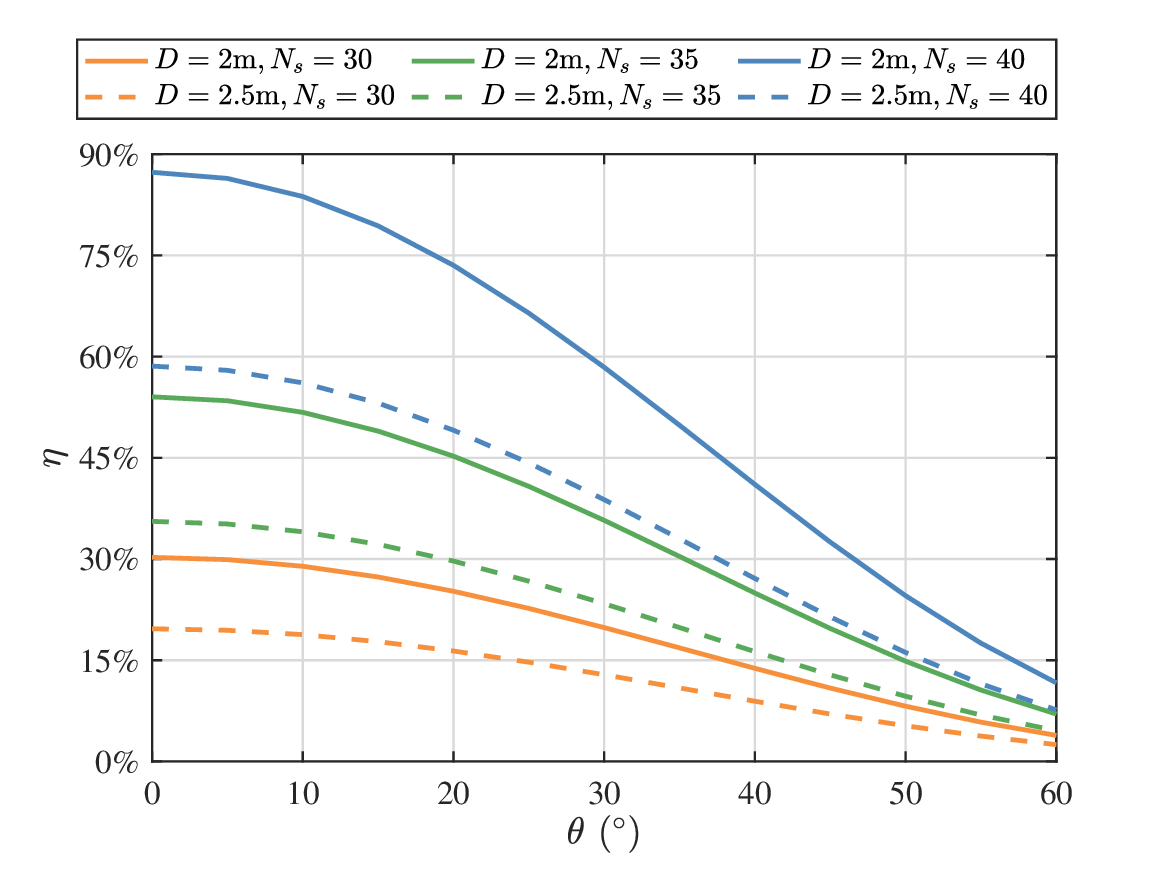}
        \caption{The impact of $\theta$ on transmission efficiency where the $\phi$=$15^\circ$}.
	\label{efficiency_theta}
\end{figure}

Fig.~\ref{efficiency_theta} illustrates the impact of various elevation angles on the radiated power at specific distances when $\phi=15^\circ$. It can be seen that as $\theta$ increases, the transmission efficiency gradually diminishes. This decrease is due to the increase of $\theta$ leading to a decrease in antenna gain. As shown in Figure~\ref{gain}, when the elevation angle increases, the gain of the target antenna decreases. According to eq.~(\ref{eq6}) and~(\ref{eq7}), the decrease in antenna gain leads to a reduction in the reception area and received power at the target end. The output power of the BS is not only affected by the antenna gain but also by the power amplifier. Therefore, the increase in angle has a smaller impact on the BS, resulting in the continuous downward trend observed in Fig.~\ref{efficiency_theta}.

\subsection{Analysis of Positioning Estimation Accuracy}

\begin{figure}
  \centering
\subfigure[]{
	\includegraphics[width=0.8\linewidth]{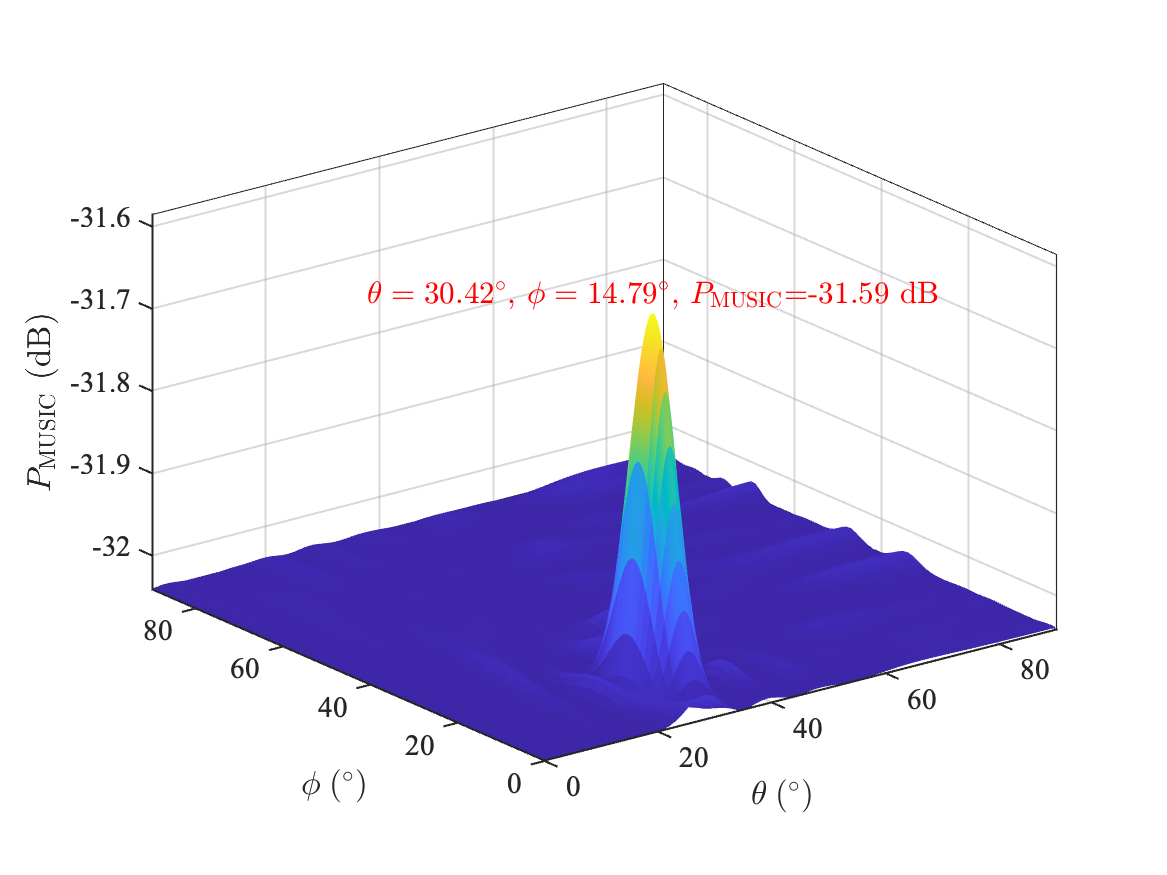}}
        \label{RBS}
\subfigure[]{
        \includegraphics[width=0.8\linewidth]{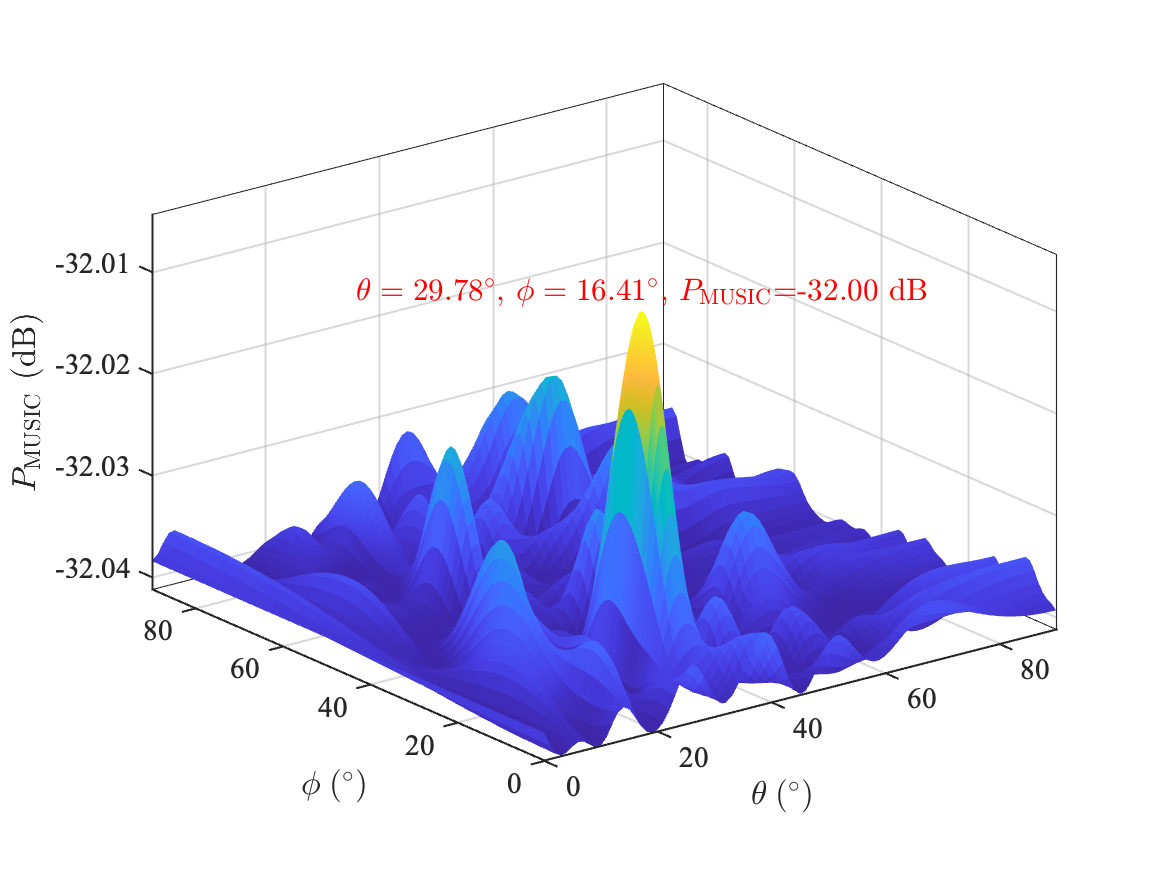}}
	\label{APS}
  \caption{3-D spatial spectrum for MUSIC algorithm under. (a) RF-RBPS, the distance $D=2.5\mathrm{m}$, the array size $N_s=40$, and the noise power $P_\mathbf{N}=0.02\mathrm{mW}$. (b) RF-APS, the distance $D=2.5\mathrm{m}$, the array size $N_s=40$, and the noise power $P_\mathbf{N}=0.02\mathrm{mW}$.}
  \label{3dmusic}
\end{figure}

In this subsection, we primarily utilize the MUSIC algorithm to assess the performance of the passive positioning system based on resonant beams as compared to the active positioning system. Fig.~\ref{3dmusic}(a) and Fig.~\ref{3dmusic}(b) display the simulation outcomes for RF-RBPS and RF-APS using a $40\times40$ array configuration, with the BS and target separated by $2.5\mathrm{m}$, and angles $\theta=30^\circ$ and $\phi=15^\circ$. It is presupposed that both systems commence with an emission power of $1\mathrm{mW}$, and the BS receives an equivalent noise power of $0.02\mathrm{mW}$.

\begin{figure}[!t]
	\centering
 \subfigure[]{
	\includegraphics[width=0.8\linewidth]{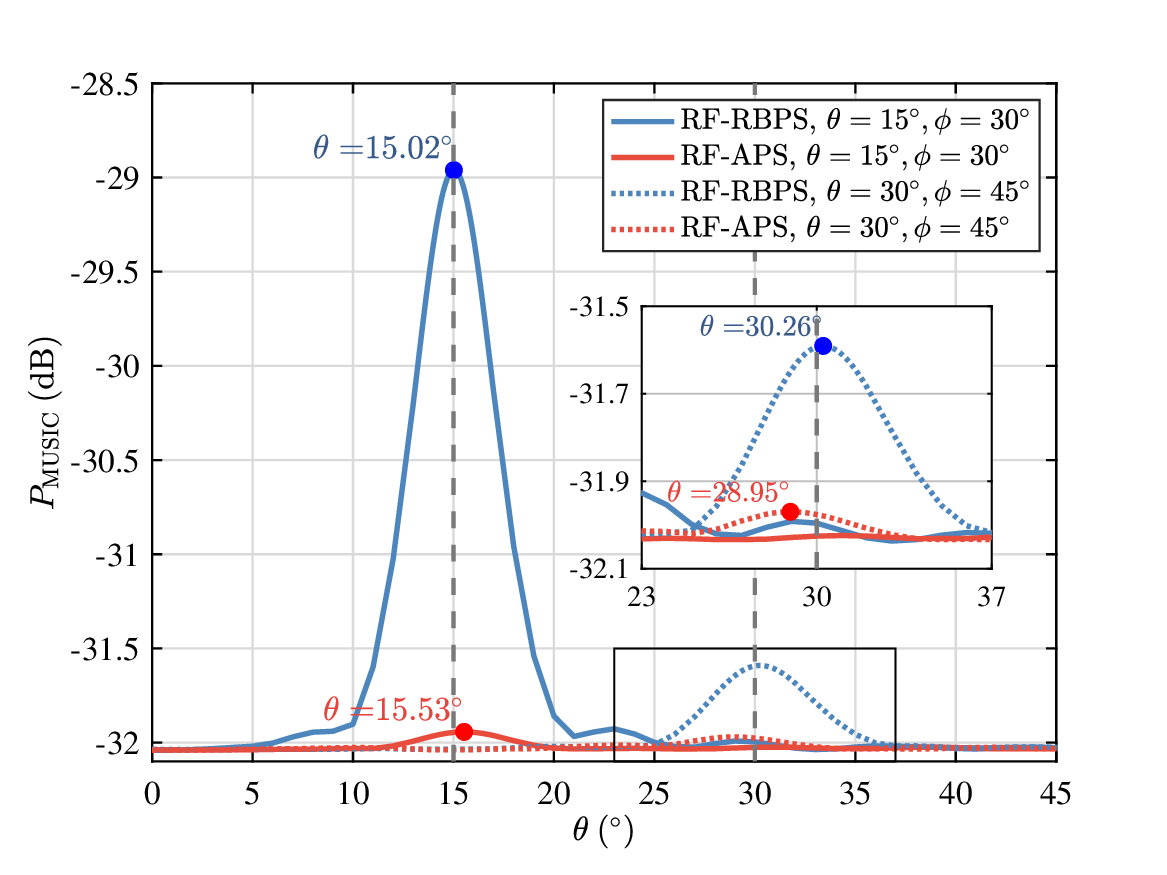}}
        \label{theta}
 \subfigure[]{
        \includegraphics[width=0.8\linewidth]{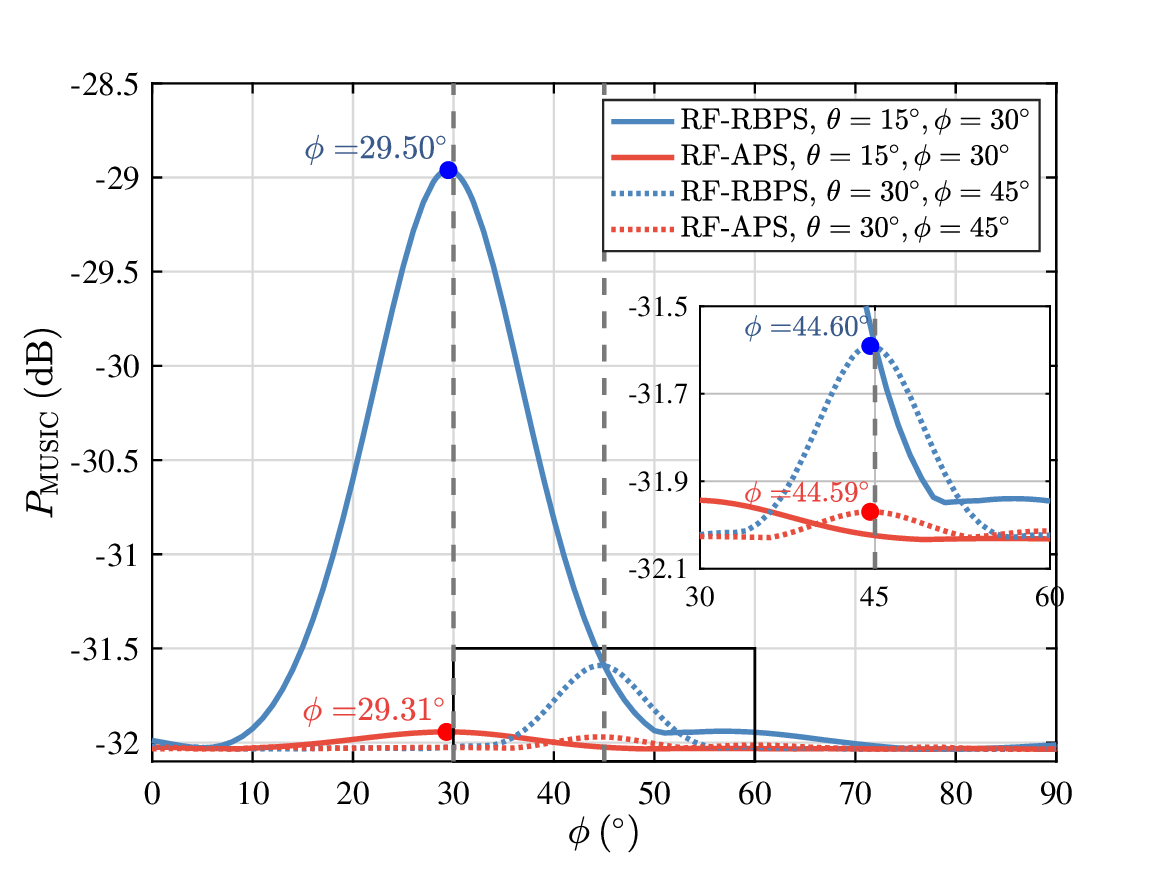}}
	\label{}
	\caption{2-D spatial spectrum for MUSIC algorithm. (a) elevation estimation, $D=2.5\mathrm{m}$, the array size $N_s=40$, and the noise power $P_\mathbf{N}=0.02\mathrm{mW}$. (b) azimuth estimation, $D=2.5\mathrm{m}$, the array size $N_s=40$, and the noise power $P_\mathbf{N}=0.02\mathrm{mW}$.}
         \label{2dmusic}
\end{figure}

The spatial spectrum computed by the MUSIC algorithm reveals that the 
spectrum of RF-RBPS is sharper and has larger peaks, indicating that the signal intensity of the signal source in Fig.~\ref{3dmusic}(a) is stronger, which is due to the difference in radiation power of the target. Moreover, it is discernible that the DoA estimates for RF-RBPS are more accurate. The primary cause for this discrepancy lies in the omnidirectional scattering and severe path loss encountered by the EM waves emitted by the target in the active positioning system. In contrast, the RF-RBPS, characterized by its energy concentration and self-alignment attributes, secures a higher power at the BS under identical initial emission power conditions, thereby facilitating more precise DoA estimation and exhibiting smoother spectral peaks in the spatial spectrum.

Fig.~\ref{2dmusic}(a) and Fig.~\ref{2dmusic}(b) provide a more detailed depiction of two sets of angles $\theta=15^{\circ}$, $\phi=30^{\circ}$ and $\theta=30^{\circ}, \phi=45^{\circ}$, and the gray dashed represents the accurate value. It is evident that the DoA accuracy of RF-APS is inferior, with its MUSIC spatial spectrum peaks being less pronounced compared to RF-RBPS. This further substantiates our proposed positioning system based on resonant beam technology, which not only accomplishes passive positioning but also surpasses the active array in terms of positioning performance. Additionally, it is observed that an increase in $\theta$ leads to a reduction in accuracy. Taking RF-RBPS as an example, as shown as Fig.~\ref{efficiency_theta}, the increased $\theta$ results in a diminished antenna gain, thereby causing a decline in transmission efficiency. Consequently, under a fixed transmission power at the BS, the reflected power at the target is reduced, hence diminishing the precision of localization.

\begin{figure}[!t]
	\centering
 \subfigure[]{
	\includegraphics[width=0.8\linewidth]{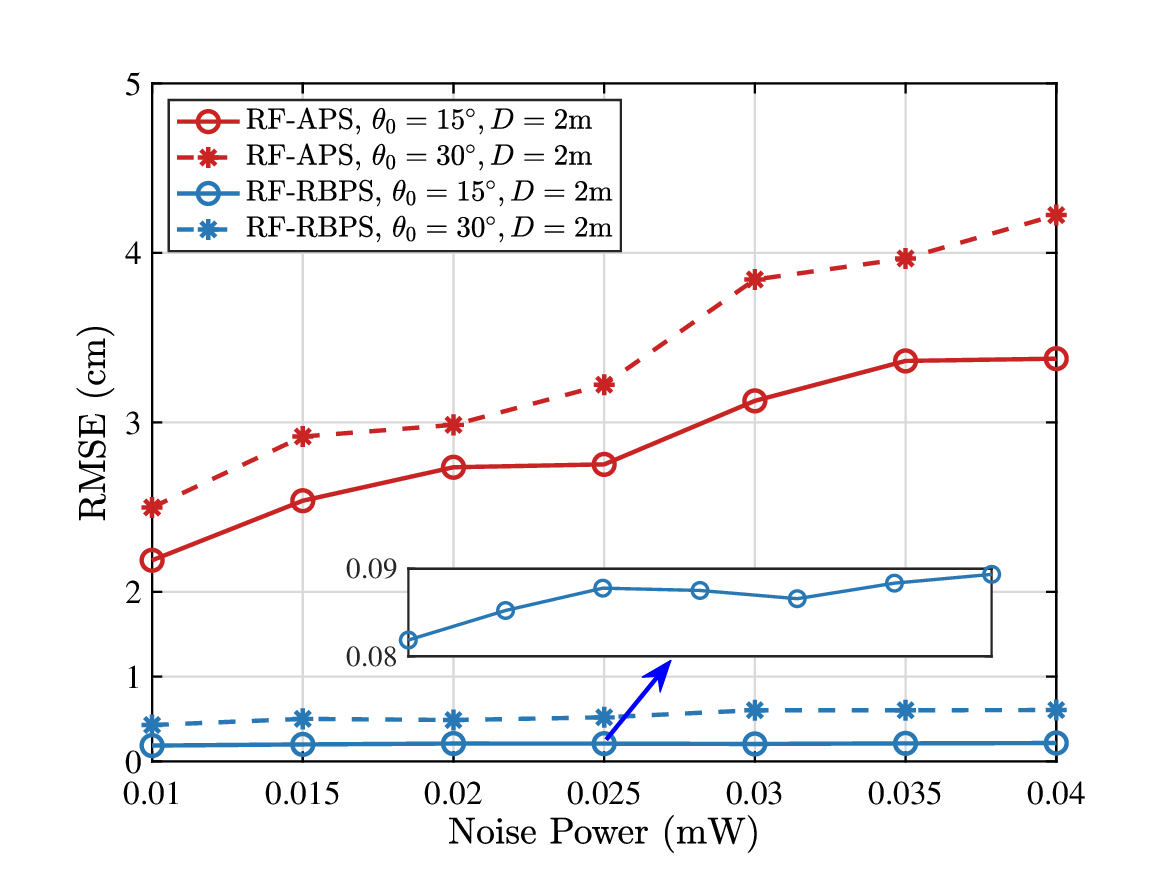}}
        \label{2m_noise}
 \subfigure[]{
        \includegraphics[width=0.8\linewidth]{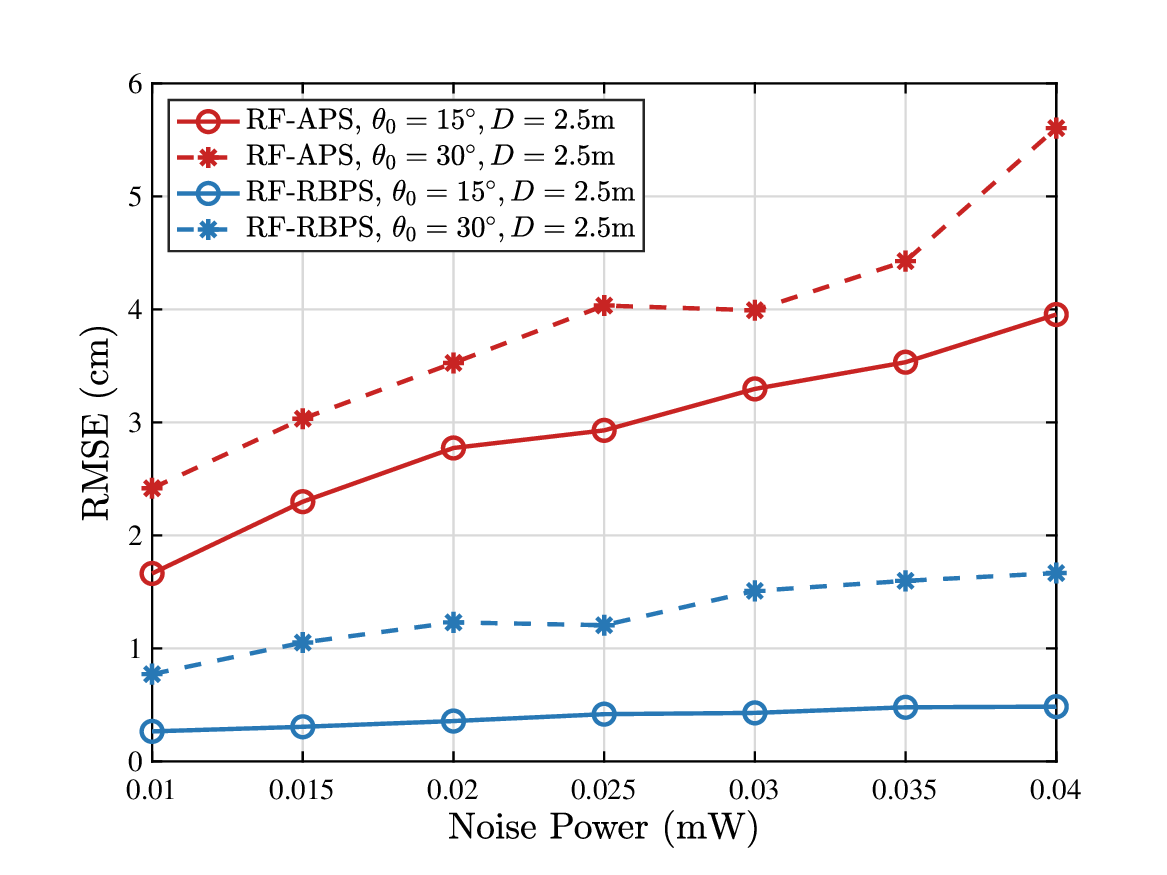}}
	\label{2.5m_noise}
	\caption{RMSE of 3-D localization with different noise powers. (a) $D=2\mathrm{m}$, (b) $D=2.5\mathrm{m}$}
         \label{rmse_noise}
\end{figure}

\begin{figure}[!t]
	\centering
 \subfigure[]{
	\includegraphics[width=0.8\linewidth]{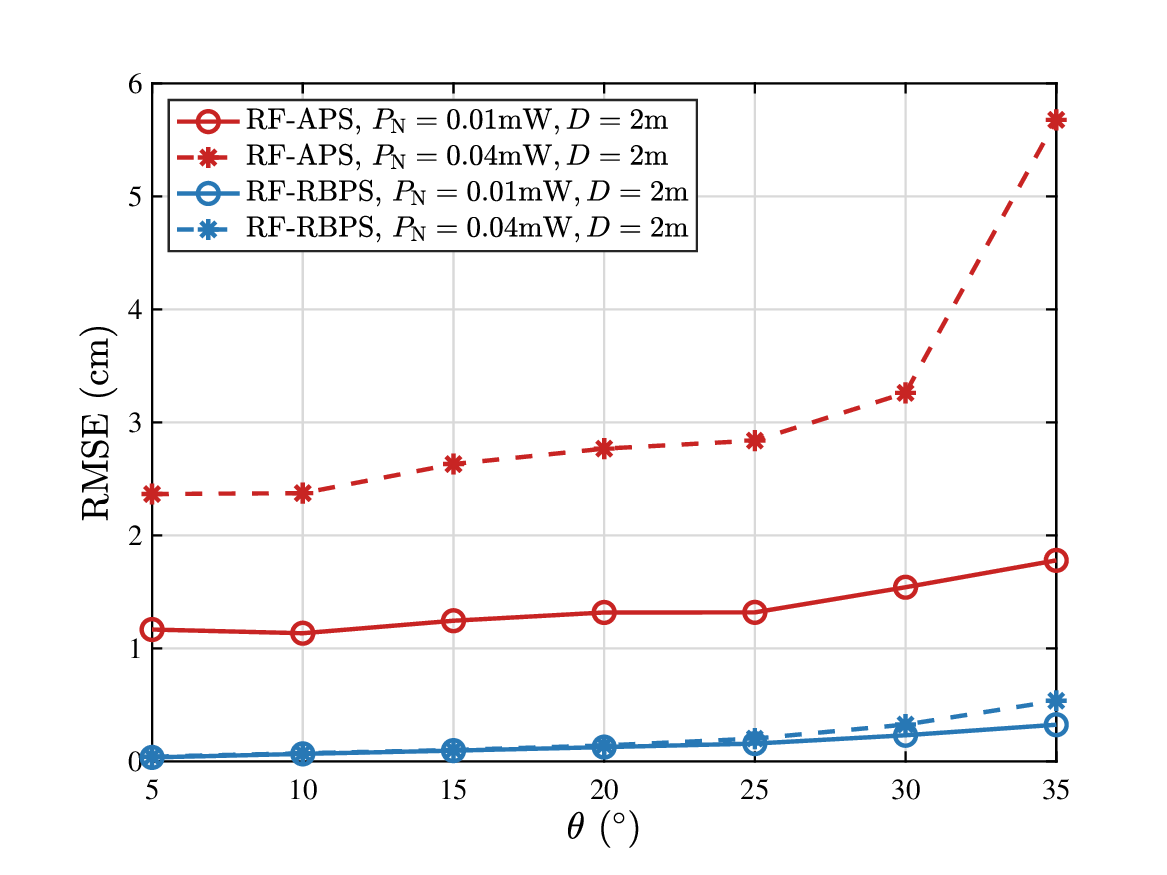}}
        \label{2m_theta}
 \subfigure[]{
        \includegraphics[width=0.8\linewidth]{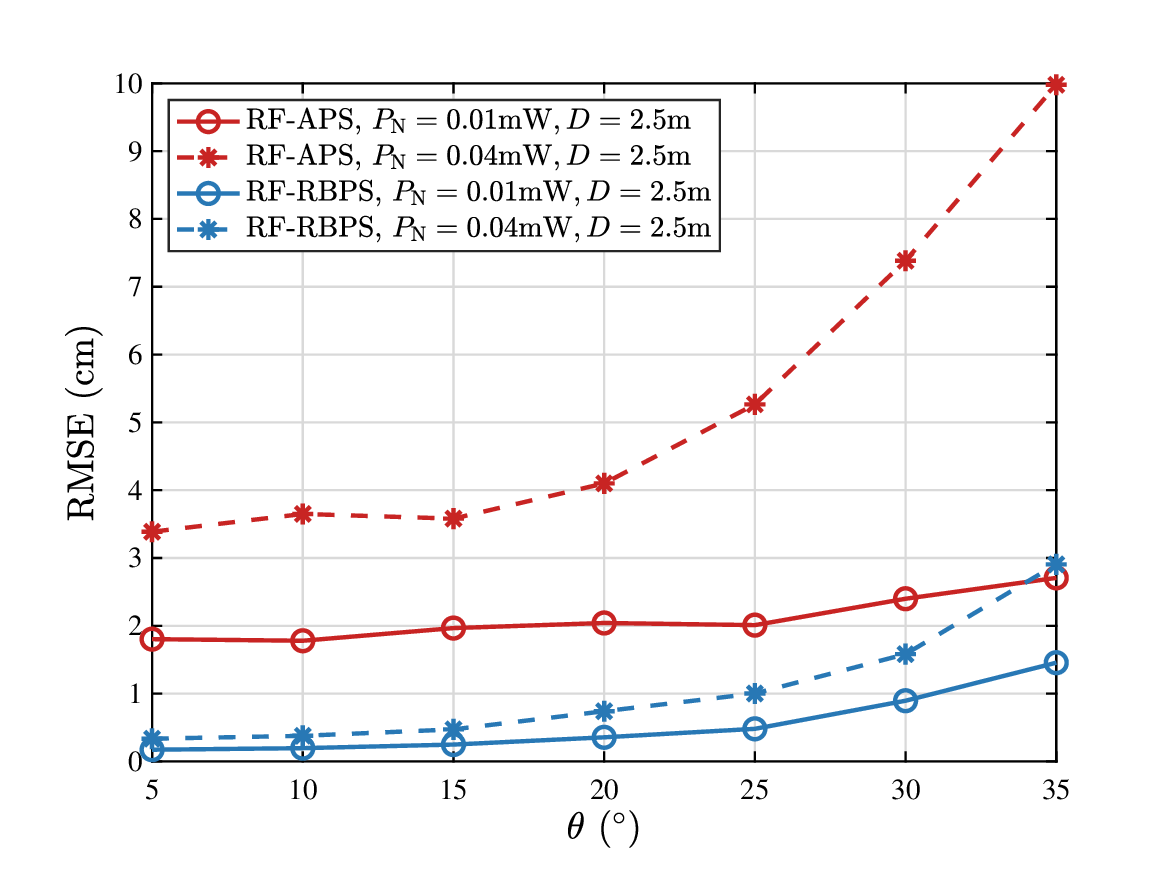}}
	\label{2.5m_theta}
	\caption{RMSE of 3-D localization with different elevation angles. (a) $D=2\mathrm{m}$, (b) $D=2.5\mathrm{m}$}
          \label{rmse_theta}
\end{figure}
To facilitate the analysis of positioning errors in both systems, Fig.~\ref{rmse_noise} and Fig.~\ref{rmse_theta} present the outcomes of $100$ Monte Carlo simulations, simulating scenarios where both the BS and target consist of $40\times40$ antenna arrays. Subsequently, the three-dimensional coordinates of the target are computed based on the angles obtained, and the RMSE is calculated by eq.~(\ref{eq29}).

Fig.~\ref{rmse_noise} illustrates the relationship between position estimation and noise power, with Fig.~\ref{rmse_noise}(a) and Fig.~\ref{rmse_noise}(b) corresponding to scenarios at $D=2\mathrm{m}$, $D=2.5\mathrm{m}$, respectively. It is observed that, for a given system, the RMSE increases with the rise in noise power. This increment is attributable to noise interference with signal reception, which impairs the signal processing algorithm’s ability to estimate the position of the signal source. Under the same noise conditions, the accuracy of RF-RBPS is less affected by noise compared to RF-APS. This improvement is due to the excellent energy concentration characteristics of RF-RBPS, which can enhance the SNR by increasing the radiated signal power from the target. Moreover, the distance between the station and target has proven to be a significant factor affecting RMSE, as larger distances exacerbate errors due to propagation effects in beamforming.

The two figures in Fig.~\ref{rmse_theta} elucidate the relationship between the RMSE of position estimation for RF-APS and RF-RBPS and the elevation angle $\theta$. Fig.~\ref{rmse_theta}(a) and Fig.~\ref{rmse_theta}(b) are for scenarios at $D=2\mathrm{m}$ and $D=2.5\mathrm{m}$, respectively. It can be inferred that, for a given system, RMSE tends to increase with the elevation angle $\theta$. This is again because the angle of signal reception becomes steeper, potentially reducing the effective receiving area of the antenna and thus lowering signal quality, which in turn diminishes the precision of the positioning algorithm. 
Compared with RF-APS, RF-RBPS has higher positioning accuracy and is less affected by elevation angle. This is because RF-RBPS has the characteristic of self-alignment. Even if the PT and BS are offset relatively large, stable cyclic power can still be formed. When the angle increases to the point where the power amplifier of the BS cannot compensate for propagation loss, the positioning accuracy will experience significant fluctuations. In RF-APS, the target emits omnidirectional radiation signals, and the BS locates by receiving partial signals from the PT. When the angle is slightly offset, the power received by the BS will be greatly reduced. Therefore, in the presence of the same noise, the SNR is lower, and the angle estimation range is smaller. Similarly,
distance is also a crucial factor impacting RMSE.

\subsection{Summary}
The simulation results show that the RF-RBPS exhibits efficient energy transmission over long distances without the need for channel estimation and beamforming, making it particularly suitable for high-precision positioning applications. The analysis of radiated power demonstrates that the electromagnetic waves undergo multiple round trips through the conjugate circuit, during which their amplitude and phase are continuously optimized, eventually reaching equilibrium. At this point, the power density and transmission efficiency stabilize. This iterative process ensures maximized transmission efficiency for the proposed system.

Furthermore, the analysis of positioning accuracy indicates that the RF-RBPS maintains high positioning accuracy under various environmental conditions, including different elevation angles and noise levels. This is attributed to the proposed system's energy concentration and self-alignment characteristics, which significantly enhance signal strength and improve the signal-to-noise ratio, facilitating more accurate position estimation even in complex scenarios.

\section{Conclusion}

In this paper, we confirm that RF-RBPS can achieve high-precision passive positioning. Firstly, we introduced a passive positioning scheme leveraging the principles of resonant beam propagation, and proposed a passive positioning system with echo-enhanced characteristics that enable bidirectional wave propagation between the BS and the PT. We estimated the DoA of the round-trip resonant beam detected at the BS end, by employing the MUSIC algorithm. Compared to the RF-APS, the RF-RBPS achieves higher accuracy in angle estimation and position calculation. Finally, simulation has confirmed that RF-RBPS can achieve high energy concentration and beam self-alignment without the need for complex channel estimation and beam control. It can also achieve millimeter level accuracy in positioning within 2m, with an error of less than 3cm within 2.5m, which is superior to active positioning systems under the same settings.

In addition, future research will focus on: 1) near-field positioning schemes; 2) Integrated design of communication positioning; And 3) Multi target localization and recognition.

\bibliographystyle{IEEEtran}

\bibliography{Mybib}

\newpage

\vfill

\end{document}